\documentclass[twocolumn]{aastex62}

\usepackage{multirow}

\usepackage{savesym}
\savesymbol{tablenum}
\usepackage{siunitx}
\restoresymbol{SIX}{tablenum}

\DeclareSIUnit\au{AU}
\DeclareSIUnit\mearth{M_\oplus}
\DeclareSIUnit\myr{Myr}

\defcitealias{2012ApJStewart}{SL12}

\received{2018 November 15}
\revised{2019 February 8}
\accepted{2019 March 2}
\submitjournal{ApJ}

\shorttitle{Fate of the runner}
\shortauthors{Emsenhuber and Asphaug}

\begin{document}

\title{Fate of the runner in hit-and-run collisions}

\correspondingauthor{Alexandre Emsenhuber}
\email{emsenhuber@lpl.arizona.edu}

\author[0000-0002-8811-1914]{Alexandre Emsenhuber}
\affil{Lunar and Planetary Laboratory, University of Arizona \\
1629 E. University Blvd. \\
Tucson, AZ 85721, United States}

\author[0000-0003-1002-2038]{Erik Asphaug}
\affil{Lunar and Planetary Laboratory, University of Arizona \\
1629 E. University Blvd. \\
Tucson, AZ 85721, United States}

\begin{abstract}

In similar-sized planetary collisions, a significant part of the impactor often misses the target and continues downrange. We follow the dynamical evolution of ``runners'' from giant impacts to determine their ultimate fate. Surprisingly, runners re-impact their target planets only about half of the time, for realistic collisional and dynamical scenarios. Otherwise they remain in orbit for tens of millions of years (the limit of our \textit{N}-body calculations) and longer, or sometimes collide with a different planet than the first one. When the runner does return to collide again with the same arget planet, its impact velocity is mainly constrained by the outcome of the prior collision. Impact angle and orientation, however, are unconstrained by the prior collision.

\end{abstract}


\section{Introduction}
\label{sec:intro}

The late stage of terrestrial planet formation is dominated by collisions at near-escaping velocities between similar-sized planetary bodies, a.k.a. giant impacts \citep{1985ScienceWetherill,2002ApJKokubo}. But giant impacts are seldom efficient when it comes to accretion \citep{2004ApJAgnor}; more often than not, a significant part of the smaller of the two (the impactor) ``misses'' the larger target. These ``hit and run collisions'' \citep{2006NatureAsphaug} can result in multiple escaping remnants or a single ``runner'', the impactor stripped of its exterior materials \citep{2009ApJMarcus}. As discussed below, and in \citet{2006NatureAsphaug} and Gabriel et al. (subm.), the basic reason is that a portion of the impactor ``misses'' the target, in the sense that their projected pathways to not fully overlap. For example, consider two differentiated planets, each with a cores that is half the planet's radius. If they impact at only \SI{30}{\degree}, more head-on than most impacts, then their cores will fly right past one another.

The fate of the runner turns out to be a nuanced and important problem. Shock dissipation and momentum transfer to the target decrease the relative velocity of the runner compared to its pre-impact velocity, relative to the target. It is helpful (but overly simplistic) to think of the runner as ``bouncing'' off the target with some damping. Depending on incoming velocity, there is a substantial deflection of the impact plane and change in trajectory during the non-accretionary close encounter.

The impactor always hits the target at faster than the mutual escape velocity, having non-zero original relative motion. However the runner -- the remnant impactor -- emerges slower than the impact speed.  Depending on dissipation, the emergence velocity can be either slower or faster than the escape velocity. When it emerges slower than the escape velocity it remains gravitationally bound, and the result is one or more successive collisions separated a day or more, in the case of a graze and merge collisions \citep[GMC;][]{2010ApJLeinhardt}. When the runner escapes the target, we call the result a hit and run collision (HRC). Explored here is the case where the escaping runner, orbiting the Sun, returns to the target on a time scale of thousands to millions of years~\citep{2012MNRASJackson}, what we call a hit and run return collision (HRR).

Despite the great variety of giant impact outcomes, and the known inefficiency of giant impact accretion, most planetary system formation models still rely on collisions being treated as perfect mergers unconditionally, with a few exceptions, e.g., \citet{1998IcarusAlexander,2013IcarusChambers}. The underlying rationale for the assumption of perfect merger, is that even if a giant impact creates multiple bodies (a target and a runner) those will remain on crossing orbits and thus will collide again at some point, resulting in total accretion in the end. This seems reasonable enough, but warrants further investigation if there are other bodies to perturb the trajectories. In addition, including multiple collision fragments increases the computational requirements, making the simulation very expensive, or untenable.

An improved model treating large collisions more realistically has been used in \citet{2015IcarBonsor,2015ApJLeinhardt,2015ApJCarter} to model the intermediate stage of terrestrial planet formation; significantly it was found that a more realistic treatment of collisions would, as expected, increase the compositional diversity of  remainders. The realistic treatment of giant impacts is also significant to the formation and evolution of gas giants and their H/He envelopes \citep{2015IcarusSchlichting}. Gas giants form while the nebula is still present, during the first 3--\SI{6}{\myr} \citep{2009AIPCMamajek,2016ApJLi}, whereas  terrestrial planet formation finishes over the next 100--\SI{200}{\myr} \citep[e.g.,][]{1999IcarusAgnor,2001IcarusChambers}. Depending on the timing and the mass distribution of giant impact remnants, the resulting bombardment by unfinished remnants can cause the loss of a giant planet envelope \citep{2016ApJInamdar}. 

Runners from HRC are mostly composed of the projectile, and have a lower mass than the projectile. In characteristic HRC the outcome is a projectile interior, stripped of its outer layers \citep{2009ApJMarcus}. This outcome could be able to explain several features of the Solar System, such as Mercury's anomalously high bulk density \citep{1988IcarusBenz,2007BenzSSRv,2014NatGeoAsphaug,2018ApJChau} and removal of water from icy planetesimals \citep{2018CeMDABurger}. Material exchange between the target and impactor during a HRC is also able to provide a means for material equilibration, where surviving runners can have signatures closer to their previous targets if they survive until today. Also, in the case of disrupted runners, surviving fragments can be genetically related, part of one original body, a kind of early catastrophic disruption that can occur at low relative velocity \citep{2017BookAsphaug}. 

The aim of this work is to properly determine the destination of the runner in HRC. If it comes right back in a slow effective merger, then it might not be overall that different from a GMC. If the runner comes back on a much longer timescale, and with randomized orientation, or does not come back at all, the consequences are of course different, and important. We model several such events with a hydrodynamical code to obtain realistic end state of HRC, and track the remnants until those collide again. In the case of re-collision we determine the parameter of the second event, and its type, and search for any correlation (or lack thereof) with the first collision. This will provide clues whether perfect accretion is a justifiable simplification to accretion codes, and will determine the time that is ultimately needed for accretion between a pair or planets to play out in orbit.

\section{Methods}
\label{sec:methods}

We adopt a coupled approach, where the initial collision is modelled with a hydrodynamical scheme,  which captures the shock process and tidal interactions. Once the bodies are separated enough so that neither plays a significant role anymore, the result are then transferred into an N-body code to obtain the dynamical evolution of the resulting bodies. The N-body evolution is performed many times for each initial collision assuming different impact orientations to obtain a statistical description.

\subsection{Collision stage}

Collisions are modelled using the Smoothed Particle Hydrodynamics (SPH) technique using a code especially suited for large scale collisions \citep{2012IcarusReufer,2014NatGeoAsphaug,2018IcarusEmsenhuber}. SPH is a Lagrangian method with material represented by particles. Quantities are retrieved by performing a kernel interpolation and spatial derivatives by interpolation of the underlying quantity with the kernel derivative \citep[see e.g.][for reviews]{1992ARA&AMonaghan,2009RosswogNAR}. We use the M-ANEOS equation of state \citep{ANEOS,2007M&PSMelosh} to obtain the pressure $p(\rho,u)$ and other necessary quantities for the hydrodynamical equations. In addition self-gravity is included by the mean a hierarchical spatial tree \citep{1986NatureBarnesHut}.

We model grazing collisions (HRC and GMC) and study the spatial mixing of particles to ascertain the effectiveness of successive giant impacts, especially on isotopic equilibration. This kind of collisions account for the vast majority of giant impacts. For collisions where the runner is not re-accreted on a short time (a couple of days), we determine in addition the spatial deflections that follow from the collision, the post-trajectory of the runner versus the pre-trajectory of the projectile. Because of the complex mass distribution and states of rotation, it sometimes is not obvious, until quite late in the evolution, that the collision will be a GMC and not a HRC. In the case of studying HRR where the runner escapes the target, the calculation of HRR has to advance the runner as an independent planet for thousands or millions of years, using dynamical methods as described below.

The initial bodies are obtained by evolving initial spheres with particles uniformly spaced, and with reference density from the equation of state, under self-gravity and pressure forces, applying a damping term and constant entropy until hydrostatic equilibrium is reached. The bodies are not spinning. Resolution is selected so that an \SI{1}{\mearth} body with a chondritic composition, that is, 30wt\% iron core and 70wt\% silicate mantle, is represented by 500,000 particles. Collision modelling is performed in the two-body center of mass frame, beginning with the bodies several radii away, as this allows for tidal deformation and spin-up prior to the contact. These effects can strongly influence the outcome, compared to starting out the targets at the time of contact. 

Collisions are first evolved to \SI{24}{\hour} after initial contact. At this point we compute the resulting properties of each body, including their orbital parameters. It is not trivial to determine the final independent bodies (e.g. target, runner) at this relatively early time, because there are spurious effects due to the fast rotation of the runner that can cause a large fraction of the runner particles (those rotating towards the target) to appear energetically bound to the target even though the two bodies are escaping from one another. The search for post-collision independent bodies is therefore performed starting with a friends-of-friends search (FoF), followed by a determination of gravitational binding. FoF clumps are treated as a single super particle during the latter stage. See appendix~\ref{sec:bodies} for details and the necessity of such an approach. The choice of ending the hydrodynamical simulation at \SI{24}{\hour} is to ensure that the body properties have converged. In all the simulations presented hereafter, the resulting properties do not change by more than a few percents after \SI{18}{\hour} (an example for mass convergence is provided in appendix~\ref{sec:bodies}).

In order to ease the comparison with the parameters of the return collision, we introduce two new quantities, $v_\mathrm{dep}$ and $\theta_\mathrm{dep}$, which are the departing velocity and angle of the secondary remnants with respect to the largest, at a distance which is the sum of their radii after the collision. These would be identical to the returning collision velocity and angle, if energy and angular momentum were conserved. The presence of the Sun and planets makes the return collision non-conservative. In the case of HRC the escaping runner must be tracked for many orbits until its next encounter with the target; this can be a close encounter leading to escape of the runner, or a follow-on collision that we model by mapping the outcome of one hydrodynamical simulation into another. 

\subsection{Dynamical evolution}
\label{sec:methods-dyn}

\begin{figure}
\centering
\includegraphics{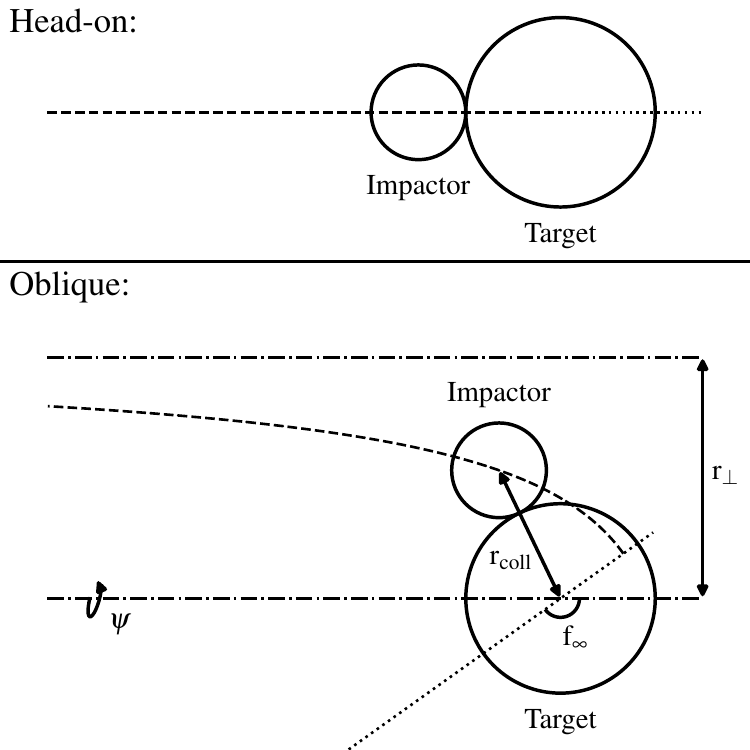}
\caption{Sketch of the correction due impact obliqueness. Note that a small mutual inclination of the orbits can cause a high inclination of the collisional plane. \textit{Top:} case of an head-on collision, where the relative distance and velocity lie on the same axis; \textit{bottom}: oblique collision with $\theta_\mathrm{coll}=\SI{45}{\degree}$ and $v_\mathrm{coll}/v_\mathrm{esc}=1.1$. The dashed lines represent the trajectory of the impactor, the dotted lines the major axis of the orbit and the dash-dotted lines the asymptote of the hyperbola, and its parallel that passes through the center of the target. At large separation, the relative position and velocity vectors tend to be parallel, with a offset $r_\bot$. }
\label{fig:orb}
\end{figure}

We map the target and runner emerging from one collision, ignoring collision products smaller than the two primary remnants, into the \textit{Mercury} \textit{N}-body code \citep{1999MNRASChambers}. The resulting bodies are followed until a subsequent collision occurs, or up to \SI{20}{\myr}. To obtain a statistics of the possible outcomes, we adopt a Monte Carlo approach assuming a range of possible pre-impact orbits. We consider the target to be on a circular orbit at \SI{1}{\au}, and the pre-impact orbit of the impactor is computed according to \citet{2018MNRASJackson}. That methodology provides the orbit for a head-on collision, as in the top panel of Fig.~\ref{fig:orb}, with the angles $\phi$ and $\theta$ providing the direction of the relative velocity vector. In the case of an oblique impact, a modification must be incorporated. We will assume that the same angles represent the direction of the asymptote of the hyperbolic orbit. We also introduce a third angle, $\psi$, that represent a rotation about this direction.

Neglecting of the distance perpendicular to the eccentricity vector introduces only a small error. The deviation from a head-on orbit at large separation is given by
\begin{equation} 
\frac{r_\bot}{r_\mathrm{coll}}=\frac{v_\mathrm{coll}/v_\mathrm{esc}}{\sqrt{(v_\mathrm{coll}/v_\mathrm{esc})^2-1}}\sin{\theta_\mathrm{coll}}
\end{equation}
with $r_\mathrm{coll}=r_\mathrm{tar}+r_\mathrm{imp}$ the separation at initial contact (this relationship follows from energy and angular momentum conservation). If we assume two Earth-size planets colliding at \SI{45}{\degree}, with a velocity of $v_\mathrm{coll}/v_\mathrm{esc}=1.1$, then the offset at large separation is $r_\bot\approx\SI{2.2e4}{\kilo\meter}$, or compared to the semi-major axis $r_\bot/a\approx\num{1.5e-4}$. This value would be the correction in eccentricity or inclination due to the obliqueness of the impact; it is sufficiently small compared to the encountered values of either quantity to be ignored. Higher impact velocities lead to even smaller corrections.

For the majority of our modeling, we assume that $\phi$ and $\theta$ have an underlying uniform distribution in space, and $\psi$ also follows a uniform distribution. We also verify the robustness of the results against specific impact geometry by comparing to a set of collision that happen only in the orbital plane. For this second series of initial conditions, we set $\theta=\pi/2$ and $\psi=0$ while $\phi$ follows an uniform distribution. Our assumption that the pre-impact orbit of the target is circular should not lead to singular situation during the dynamical evolution. The target also being deflected during the collision, its orbit will no longer be circular afterwards.

Using one ``snapshot'' of a given original collision, and rotating it in space to provide different dynamical initial states, assumes that other planets or the Sun have negligible effects on the collision process. In ongoing work (Emsenhuber \& Asphaug, in prep.) we do find that GMC of two satellites in the presence of a central planet are strongly affected by their orientation in space, but this happens because the Hill sphere is so small. The present situation is quite different. First, the Hill sphere of an Earth-mass object around the Sun is $\simeq1\%$ of its distance to the central star. At \SI{1}{\au}, the Hill sphere is $\simeq\SI{1.5e6}{\kilo\meter}$ or more than 200 Earth radii. Only the most weakly-bound GMCs would be affected by the presence of the Sun, and the region in the parameter space (impact velocity and angle) for which this happens is very narrow.
To provide an idea of whether the transient remnant of the GMC might be affected by the presence of other bodies, the apocenter is computed for the period between the initial and second collision.

\subsection{Return collision}

Dynamical evolution is used to provide the parameters of subsequent collisions in HRR scenarios, where the runner returns for a second giant impact. We do not perform direct SPH simulations of the returning collisions, however, as this would be a considerable resource requirement, and unique in every case. Rather, we opt for ascertaining the return collision regime by using scaling laws, following either \citet{2012ApJLeinhardt} or by \citet{2019ApJCambioni}. The former is a well established model, which is based on extensive hydrodynamical simulations. The latter case is an application of machine-learning supervised classification, where the underlying collisions were performed using a similar method and initial conditions as in the present work. In particular, the classes provided by the latter are consistent with the description of GMCs and HRCs used here.

\section{Results}
\label{sec:results}
We describe the results of our numerical investigations, divided into the three sequential components of HRR: the initial collisions, modeled using SPH, the dynamical evolution (with and without the presence of other planets) modeled using \textit{Mercury}, and the return collisions, classified according to scaling laws. We search for correlations between initial collision parameters and return collision parameters. Contrary to the assumptions made in previous research, for a large fraction of HRC there is no return collision, in which case we categorize the fate of the runner for up to 20 million years, whether it impacts another planet (not the original target) or remains in orbit around the Sun.

\begin{rotatetable*}
\begin{deluxetable*}{ccccccccccccccc}
    \tablecaption{Initial and final conditions for the prior collision simulations\label{tab:coll}}
    \tablehead{
        \colhead{$m_\mathrm{tar}$ [\si{\mearth}]} & \colhead{$m_\mathrm{imp}$ [\si{\mearth}]} & \colhead{$\gamma$} & \colhead{$\frac{v_\mathrm{coll}}{v_\mathrm{esc}}$} & \colhead{$\theta_\mathrm{coll}$ [\si{\degree}]} & \colhead{$m_\mathrm{lr}$ [\si{\mearth}]} & \colhead{$m_\mathrm{sr}$ [\si{\mearth}]} & \colhead{$\gamma_\mathrm{after}$} & \colhead{$m_\mathrm{lost}$ [\si{\mearth}]} & \colhead{$\frac{v_\mathrm{dep}}{v_\mathrm{esc}}$} & \colhead{$\theta_\mathrm{dep}$ [\si{\degree}]} & \colhead{$f_\mathrm{core}^\mathrm{l \leftarrow i}$} & \colhead{$f_\mathrm{core}^\mathrm{s \leftarrow t}$} & \colhead{$f_\mathrm{man}^\mathrm{l \leftarrow i}$} & \colhead{$f_\mathrm{man}^\mathrm{s \leftarrow t}$}
    }
    \startdata
    \multirow{9}{*}{0.9} & \multirow{9}{*}{0.2} & \multirow{9}{*}{0.22} & 1.10 & 52.5 & 0.90\tablenotemark{b} & 0.16\tablenotemark{b} & 0.18\tablenotemark{b} & \num{1.6e-3} & 0.99\tablenotemark{b} & 57\tablenotemark{b} & 1.1\% & 0.0\% & 3.7\% & 8.1\% \\
    & & & 1.10 & 55.0 & 0.92 & 0.18 & 0.19 & \num{8.0e-4} & 1.01 & 62 & 0.6\% & 0.0\% & 5.1\% & 8.7\% \\
    & & & 1.10 & 60.0 & 0.92 & 0.18 & 0.20 & \num{2.2e-4} & 1.03 & 61 & 0.0\% & 0.0\% & 3.6\% & 6.2\% \\
    & & & 1.15 & 45.0 & 0.90\tablenotemark{b} & 0.13\tablenotemark{b} & 0.15\tablenotemark{b} & \num{4.1e-3} & 0.98\tablenotemark{b} & 53\tablenotemark{b} & 4.1\% & 0.0\% & 4.7\% & 14.2\% \\
    & & & 1.15 & 52.5 & 0.92 & 0.18 & 0.19 & \num{1.4e-3} & 1.05 & 59 & 0.8\% & 0.0\% & 5.1\% & 10.0\% \\
    & & & 1.15 & 60.0 & 0.91 & 0.19 & 0.20 & \num{3.8e-4} & 1.08 & 63 & 0.0\% & 0.0\% & 3.3\% & 5.7\% \\
    & & & 1.20 & 42.5 & 0.95 & 0.14 & 0.14 & \num{7.2e-3} & 1.00 & 53 & 5.2\% & 0.0\% & 9.0\% & 18.9\% \\
    & & & 1.20 & 45.0 & 0.94 & 0.16 & 0.17 & \num{4.5e-3} & 1.03 & 50 & 3.5\% & 0.0\% & 7.2\% & 16.5\% \\
    & & & 1.20 & 52.5 & 0.92 & 0.18 & 0.19 & \num{2.6e-3} & 1.10 & 60 & 0.5\% & 0.0\% & 4.7\% & 8.7\% \\ \hline
    \multirow{8}{*}{0.9} & \multirow{8}{*}{0.5} & \multirow{8}{*}{0.56} & 1.10 & 45.0 & 1.38\tablenotemark{a} & -- & -- & \num{1.6e-2} & -- & -- & -- & -- & -- & -- \\
    & & & 1.10 & 60.0 & 0.92 & 0.48 & 0.53 & \num{1.0e-3} & 1.00 & 60 & 0.0\% & 0.0\% & 6.1\% & 6.9\% \\
    & & & 1.15 & 45.0 & 0.87\tablenotemark{b} & 0.44\tablenotemark{b} & 0.50\tablenotemark{b} & \num{4.3e-3} & 0.96\tablenotemark{b} & 50\tablenotemark{b} & 0.8\% & 0.0\% & 8.0\% & 11.0\% \\
    & & & 1.15 & 52.5 & 0.92 & 0.48 & 0.52 & \num{2.4e-3} & 1.03 & 56 & 0.0\% & 0.0\% & 8.0\% & 10.2\% \\
    & & & 1.15 & 60.0 & 0.91 & 0.49 & 0.54 & \num{1.1e-3} & 1.06 & 62 & 0.0\% & 0.0\% & 5.2\% & 6.4\% \\
    & & & 1.20 & 45.0 & 0.92 & 0.47 & 0.51 & \num{5.3e-3} & 1.02 & 51 & 0.6\% & 0.0\% & 10.2\% & 13.5\% \\
    & & & 1.20 & 52.5 & 0.92 & 0.48 & 0.53 & \num{2.8e-3} & 1.09 & 58 & 0.0\% & 0.0\% & 7.4\% & 9.3\% \\
    & & & 1.20 & 60.0 & 0.91 & 0.49 & 0.54 & \num{1.5e-3} & 1.12 & 63 & 0.0\% & 0.0\% & 4.6\% & 5.8\% \\
    \enddata
    \tablecomments{See main text for an explanation of the symbols.}
    \tablenotetext{a}{GMC where the second collision can be modelled with uninterrupted hydrodynamical scheme.}
    \tablenotetext{b}{GMC close to the HRC regime boundary where it is not feasible to model the secondary collision with uninterrupted hydrodynamical scheme. Analysis is based on FoF rather than gravity search.}
\end{deluxetable*}
\end{rotatetable*}

\begin{figure*}
\centering
\includegraphics{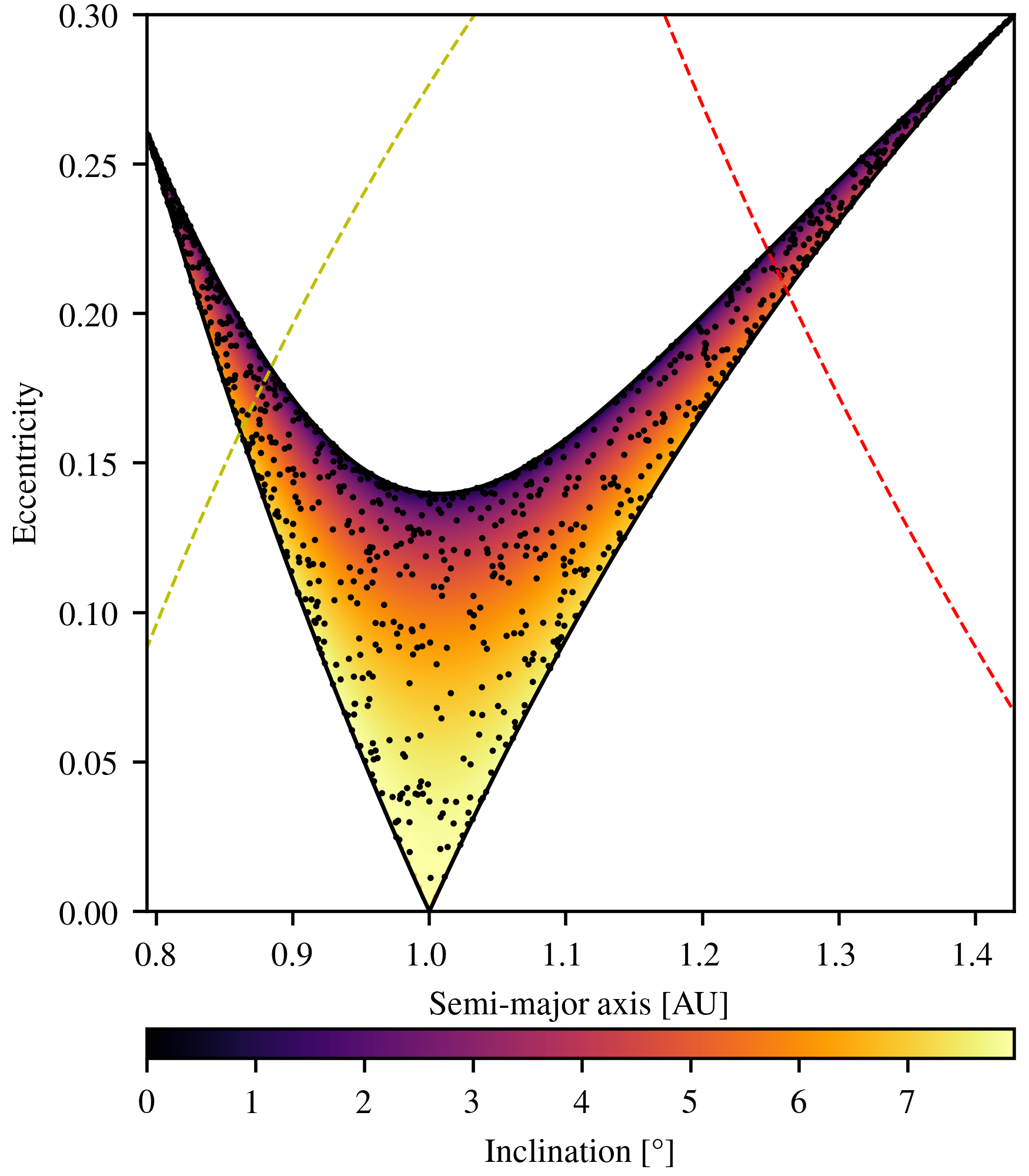}
\includegraphics{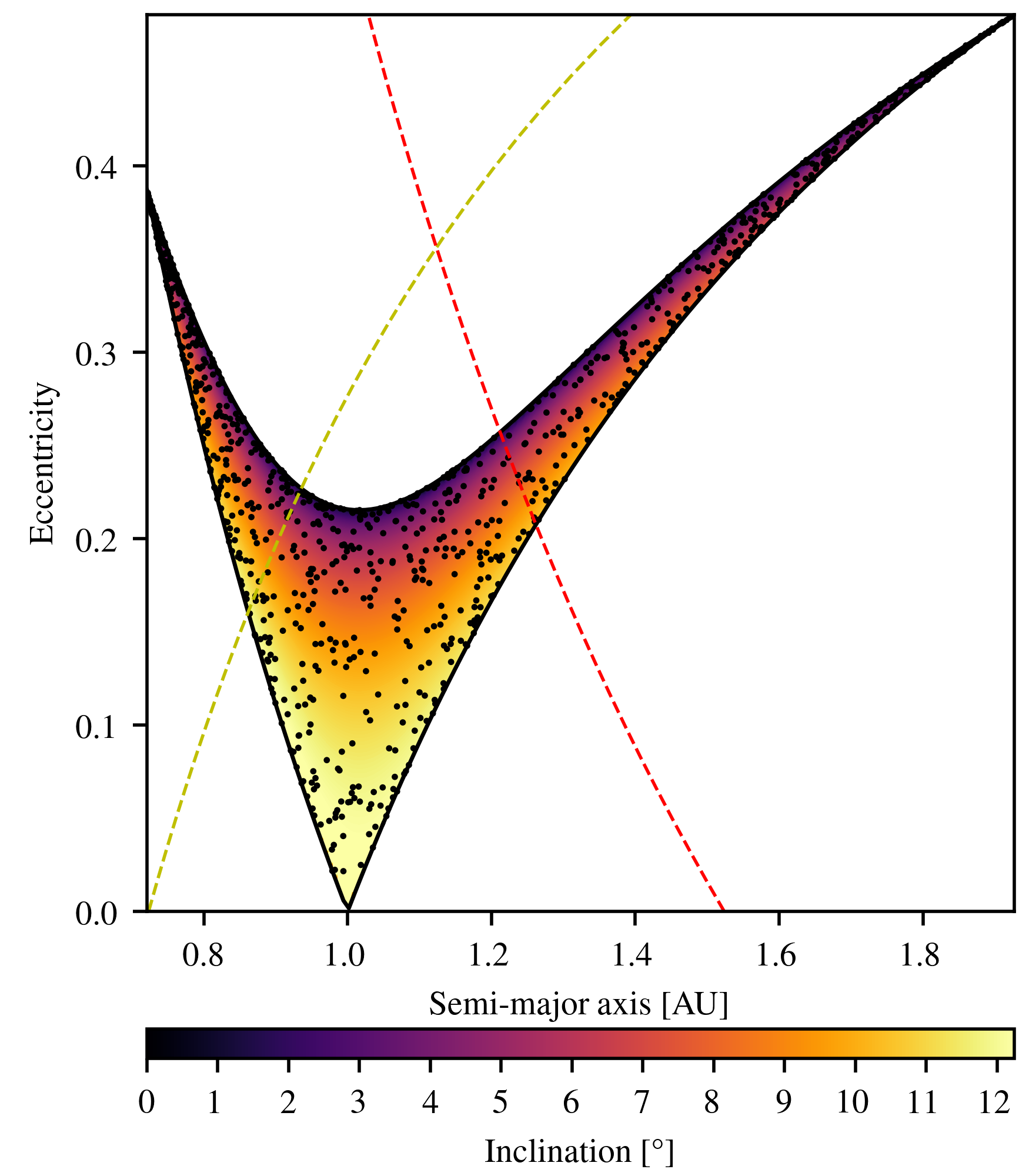}
\caption{Pre-impact orbits of the impactors in our dynamical modelling. \textit{Left}: case with $m_\mathrm{imp}=\SI{0.2}{\mearth}$ and $v_\mathrm{coll}/v_\mathrm{esc}=1.10$; \textit{Right}: $m_\mathrm{imp}=\SI{0.5}{\mearth}$ and $v_\mathrm{coll}/v_\mathrm{esc}=1.20$. These are the two extremes of our dynamical evolution, i.e. the former has the lowest relative velocity while the latter has the highest. Orbits higher than the dashed yellow curve are Venus-crossing and the ones higher than the dashed red curve are Mars-crossing.}
\label{fig:orb-pre-impact}
\end{figure*}

\subsection{Initial collisions}

We model a set of prior collisions where the mass of the target is always $m_\mathrm{tar}=\SI{0.9}{\mearth}$ and the impactor mass is either $m_\mathrm{imp}=\SI{0.2}{\mearth}$ or $m_\mathrm{imp}=\SI{0.5}{\mearth}$, so that the mass ratio $\gamma=m_\mathrm{imp}/m_\mathrm{tar}$ is 0.22 and 0.56 respectively. We limit our study to impact velocities in the range $v_\mathrm{coll}/v_\mathrm{esc}=1.1-1.2$ where $v_\mathrm{esc}=G(m_\mathrm{tar}+m_\mathrm{imp})/(r_\mathrm{tar}+r_\mathrm{imp})$ is the mutual escape velocity. The selected mass ratios fall within the expected range occurring during planetary formation. The smaller mass ratio is selected to be similar to potential Moon forming collisions \citep[e.g.][]{2012IcarusReufer}, whereas the higher one is close to the expected median of distribution during the overall formation process. More dissimilar bodies are also expected to lead to greater mass transfer from the impactor to the target during a collision, as equal-mass bodies should have none for symmetry. Different spins may break the symmetry however.

During the presence of the gas disk, our chosen velocity range is on the upper part of the distribution (Emsenhuber et al., in prep.), and close to the median found after the dispersal of the gas disk \citep{2009RaymondIcarus,2012ApJStewart}. As for impact angle, we study collision angles greater than \SI{40}{\degree} that result in grazing collisions (HRC or GMC) for these velocities. The corresponding pre-impact orbits of the impactors are shown in Fig.~\ref{fig:orb-pre-impact}. The black dots represent each initial condition for the dynamical evolution. In the case of impacts happening in the orbital plane, all the pre-impact orbits have $I=0$ and lie on the upper boundary of each plot. For the lowest velocity case, $v_\mathrm{coll}/v_\mathrm{esc}=1.1$, this means that the impactor has an eccentricity of about 0.15 (if it has the same semi-major axis than the target, more otherwise), an inclination up to about \SI{8}{\degree}, or a combination of both. The pre-impact orbits of the larger impactors are slightly more eccentric or inclined for the same impact-to-escape velocity ration, as the mutual escape velocity increases by nearly 7\%. The mass of the impactor more than doubles, but the separation at initial contact also increases, which counterbalances the effect.

The main results from the collision modeling are provided in Table~\ref{tab:coll}. The first five columns provide the initial conditions; $m_\mathrm{lr}$ denotes the mass of the largest remnant, $m_\mathrm{sr}$ the mass of the second remnant, $\gamma_\mathrm{after}$ is the mass ratio between these two bodies, and $m_\mathrm{lost}$ is the mass bound neither to the largest nor the second remnants. $v_\mathrm{dep}/v_\mathrm{esc}$ and $\theta_\mathrm{dep}$ provide the orbital configuration of the second remnant with respect to the largest. The last four columns indicate material exchange occurring during the encounter, with the superscript $\mathrm{l}\leftarrow\mathrm{i}$ indicating the fraction of the largest remnant coming from the impactor and $\mathrm{s}\leftarrow\mathrm{t}$ the fraction of the second remnant coming from the target. Some of these collisions are GMC, indicated by 1) either the almost total accretion of the impactor, with no secondary remaining and only loose material or 2) a bound secondary body, that has yet not been accreted. In the latter case, we provide the results as found by the FoF search so that the bodies can be distinguished. The only exception to this is the lost mass, which still comes from the gravity search; so that the masses shown on the table do not add up because bound material not part of a FoF body is not accounted for in any item. For these cases, we provide a few additional quantities in Table~\ref{tab:transient}, which are computed assuming a two-body problem without external perturbation. The first quantity is the pericenter given in terms of the radius of the larger body, the second is the apocenter and the last one is the orbital period. For reference the Hill radius of the Earth is roughly \SI{1.5e6}{\kilo\meter}, so that all these transient bodies would remain within the Hills sphere during the orbit. What happens to a GMC when apocenter is comparable to the Hills radius is the subject of ongoing research (Emsenhuber \& Asphaug in prep.). 
A GMC with a return period of the order of five days could potentially be evolved with the hydrodynamical scheme with enough precision; longer-period GMCs would require huge computational resources for little benefit.

Some HRC are just above the boundary of GMC, meaning that the runner leaves the target with a velocity that is just above the mutual escape velocity. In terms of the first giant impacact, the transition between GMC and HRC is smooth; the only way to discern the longest-period GMC from the lowest-energy HRC is to integrate the final bodies forward in time. As expected, the higher velocity collisions at lower impact angles liberate greater quantities of loose material owing to the greater strength of the interaction. The exception are GMCs that involve multiple underlying encounters; for example the GMC simulation with $m_\mathrm{imp}=\SI{0.5}{\mearth}$, $v_\mathrm{coll}/v_\mathrm{esc}=1.10$, $\theta_\mathrm{coll}=\SI{45}{\degree}$ has a quite higher lost mass than the other collisions, since the GMC process involves a succession of bound giant impacts, while a faster HRC collision has only a single encounter.

As the mass ratio $\gamma$ increases, both bodies become similar in size, so that for even a relatively head-on impact angle, only a minor fraction of the impactor and target physically intersect \citep{2010AsphaugChEG}. This is not the case for small $\gamma$ where for all but the most grazing angles the target blocks the entire impactor, transitioning to the cratering regime. In the specific cases considered here, assuming spheres passing through spheres, half of the impactor volume intersects the target volume for an impact angle of \SI{37}{\degree} with the \SI{0.2}{\mearth} impactor, and \SI{33}{\degree} for the \SI{0.5}{\mearth} one. These are more head-on than the average collision, yet still count as ``grazing'' due to the similar-sized geometry. The transition between grazing and non-grazing happens at approximately these angles, with a window of about \SI{10}{\degree} on steeper impact angles where GMCs still occur, with steeper angled transitioning to more direct (although, never perfect) merger.

The impact angle for which the regime change between GMC and HRC occurs, depends on the mass ratio and the velocity \citep{2010ApJKokubo}; for low-velocity collision (close to the mutual escape velocity) only the most grazing collisions result in HRC, as a small energy dissipation is sufficient to bind the bodies. On the other hand, higher velocity collisions need to dissipate more energy, requiring a steeper impact angle to be collisionally captured. In this work, we focus on collisions around \SI{45}{\degree} because they are the most common \citep{1962BookShoemaker} and because the transitions are observed to occur around this angle. 

Our results are in agreement with the general outcome from \citet{2010ApJKokubo}. There is however a small discrepancy concerning the dependence on the impact angle with the mass of the bodies. In our results, we observe a slight shift towards lower angles as the mass ratio increase for any velocity.
In our results, there are no simulations which are in different regimes in the sets with $m_\mathrm{imp}=\SI{0.2}{\mearth}$ and $m_\mathrm{imp}=\SI{0.5}{\mearth}$ for the same impact velocity and angle; however the regime change can be inferred from the departing velocity. We note that overall the departing velocity, given is terms of the escape velocity of the resulting bodies is lower for the higher mass ratio $\gamma=0.56$ than for the lower value $\gamma=0.22$ for the otherwise same initial conditions. The regime change occurs at $v_\mathrm{dep}/v_\mathrm{esc}=1$, as the difference between GMC and HRC is essentially whether the impactor's remnant is bound to the target past the first encounter. So the transition is pushed by a small amount towards lower angles for a given impact-to-escape velocity ratio as the mass of the impactor increases. This implies that HRCs are slightly less likely for more equal size bodies, independently of the impact velocity.

\begin{deluxetable*}{cccccccc}
    \tablecaption{Additional results for initial collisions that are in the GMC but where the secondary body has not yet been accreted\label{tab:transient}}
    \tablehead{
        \colhead{$m_\mathrm{tar}$ [\si{\mearth}]} & \colhead{$m_\mathrm{imp}$ [\si{\mearth}]} & \colhead{$\gamma$} & \colhead{$\frac{v_\mathrm{coll}}{v_\mathrm{esc}}$} & \colhead{$\theta_\mathrm{coll}$ [\si{\degree}]} & \colhead{$q/r_\mathrm{lr}$} & \colhead{$r_\mathrm{apo}$ [\si{\kilo\meter}]} & \colhead{$T$ [\si{\hour}]}
    }
    \startdata
    0.9 & 0.2 & 0.22 & 1.10 & 52.5 & 1.22 & \num{5.87e5} & 436 \\
    0.9 & 0.2 & 0.22 & 1.15 & 45.0 & 1.15 & \num{2.45e5} & 122 \\
    0.9 & 0.5 & 0.56 & 1.15 & 45.0 & 1.15 & \num{1.74e5} & 344 \\
    \enddata
    \tablecomments{The first five columns are the initial conditions. $q/r_\mathrm{lr}$ is the pericenter, given in terms of the radius of the largest remnant, $r_\mathrm{apo}$ is the apocenter and $T$ is the orbital period.}
\end{deluxetable*}

\subsubsection{Material transfer during collision}

HRCs lead to material exchange, with a net result of mass transfer from the impactor to target, and sometimes vice-versa. In one case we obtain a runner whose mantle is composed of 18.9\%  target material. Mass transfer from projectile to the target can also be more substantial, but is diluted by the mass ratio of the collision. We find that up to 9.0\% of the final target's mantle can be composed of impactor material in the $m_\mathrm{imp}=\SI{0.2}{\mearth}$ collisions, and for the more massive targets $m_\mathrm{imp}=\SI{0.5}{\mearth}$, they can end up with as much as 10.2\% impactor material. So after one HRC the remnants have a cosmochemical correlation at the ten-percent level that they did not share before the collision. Mass transfer is generally only important for collisions at steep angles, more head-on, as these lead to the greatest interactions between the bodies. Only high-velocity collisions are able to produce HRC at steep angles, as otherwise they fall into the GMC regime. Low-velocity HRCs are limited to shallow impact angles, where only minimal material interaction occurs, and are therefore unable to provide significant mixing.

We note a smaller effect on the metallic cores. Since cores are deep down in the body, they are more difficult to reach by the shock, and require steeper angles for any material interaction. As such it nearly impossible to extract target core material, and we have not identified any case where target core material is transferred to the runner. The opposite however does occur, where the impactor is so thoroughly shredded by the target that some of its core stays behind. In the simulations with $m_\mathrm{imp}=\SI{0.2}{\mearth}$, the target accreted up to 5.2\% of the impactor's core. For the cases without any transfer, both were for the most grazing collisions we modeled in this study, $\theta_\mathrm{coll}=\SI{60}{\degree}$. For the simulations with $m_\mathrm{imp}=\SI{0.5}{\mearth}$, only two show some transfer, both with $\theta_\mathrm{coll}=\SI{45}{\degree}$, with a values of less than 1\%. The small mass ratio in the former case makes it easier for the target to extract impactor's core compared to the latter case. This implies that there is an optimum mass ratio for core transfer onto the target. We estimate that the lower-mass impactors are close to the most efficient location, hence target's core would not be diluted by more than 10\% in the best case scenario. Higher-velocity impacts allow for more head-on HRC, which have the potential to allow more impactor core to be transferred onto the target during such an event.

The most basic net effect is that the mass ratio of the resulting bodies $\gamma_\mathrm{after}$ is lower than the pre-collision value $\gamma$. As most of material transfer involves mantle material, the core mass fraction of the runner is increased. This effect has been invoked to explain Mercury's anomalous high density \citep{2014NatGeoAsphaug}, where successive collisions would leave a runner highly depleted of its mantle. As mantle stripping of the runner requires a high mass contrast, such a HRC would have minor effect on the core mass fraction of the target, primarily by contributing stripped impactor silicate onto the target, both right away (as modeled here) and as a longer sweep-up \citep{2017BookAsphaug}.

\subsubsection{Mass loss}

We track only the largest and second remnant of each collision. That is because the collisions involved in this work are still in the low velocity regime so that no fragmentation occurs, just mass loss in the form of stripping and shedding of unresolved materials. The mass of the tertiary remnants, $m_\mathrm{lost}$, that we neglect is always less than 1\% of the total mass involved, for the collisions studied here. We note that the GMCs usually eject a higher mass fraction of sparse material, than the HRCs. 
This is because the ejected material carries a part of the angular momentum away, which in case of GMC must be released as the main body cannot sustain the resulting spin \citep[e.g.,][]{2013IcarusAsphaug}. In HRC, the runner carries away a significant part of the angular momentum and therefore there is no need to eject material for this purpose. Since mass loss is small and the ejection velocity is smaller than the impact velocity, the resulting equivalent of the impact angle $\theta_\mathrm{dep}$ must increase to achieve angular momentum conservation.

\subsection{Dynamical evolution}

\subsubsection{Destination}

\begin{deluxetable*}{ccccccccccc}
    \tablecaption{Fractions of destination of the runner after \SI{20}{Myr} of dynamical evolution.\label{tab:return-frac}}
    \tablehead{
        \colhead{$m_\mathrm{tar}$ [\si{\mearth}]} & \colhead{$m_\mathrm{imp}$ [\si{\mearth}]} & \colhead{$\gamma$} & \colhead{$v_\mathrm{coll}/v_\mathrm{esc}$} & \colhead{$\theta_\mathrm{coll}$ [\si{\degree}]} & \colhead{$f_\mathrm{Tar}$} & \colhead{$f_\mathrm{Mercury}$} & \colhead{$f_\mathrm{Venus}$} & \colhead{$f_\mathrm{Mars}$} & \colhead{$f_\mathrm{Other}$} & \colhead{$f_\mathrm{Rem}$}
    }
    \startdata
    \multirow{14}{*}{0.9} & \multirow{14}{*}{0.2} & \multirow{14}{*}{0.22} & \multirow{2}{*}{1.10} & \multirow{2}{*}{55.0} & 0.558 & 0.012 & 0.188 & 0.016 & 0.067 & 0.159 \\
    & & & & & 0.953 & -- & -- & -- & -- & 0.047 \\
    & & & \multirow{2}{*}{1.10} & \multirow{2}{*}{60.0} & 0.442 & 0.009 & 0.238 & 0.015 & 0.092 & 0.204 \\
    & & & & & 0.951 & -- & -- & -- & -- & 0.049 \\
    & & & \multirow{2}{*}{1.15} & \multirow{2}{*}{52.5} & 0.360 & 0.018 & 0.306 & 0.015 & 0.087 & 0.214 \\
    & & & & & 0.936 & -- & -- & -- & -- & 0.064 \\
    & & & \multirow{2}{*}{1.15} & \multirow{2}{*}{60.0} & 0.325 & 0.025 & 0.259 & 0.037 & 0.120 & 0.234 \\
    & & & & & 0.904 & -- & -- & -- & -- & 0.096 \\
    & & & \multirow{2}{*}{1.20} & \multirow{2}{*}{42.5} & 0.640 & 0.011 & 0.158 & 0.011 & 0.042 & 0.138 \\
    & & & & & 0.968 & -- & -- & -- & -- & 0.032 \\
    & & & \multirow{2}{*}{1.20} & \multirow{2}{*}{45.0} & 0.405 & 0.014 & 0.285 & 0.016 & 0.078 & 0.202 \\
    & & & & & 0.937 & -- & -- & -- & -- & 0.063 \\
    & & & \multirow{2}{*}{1.20} & \multirow{2}{*}{52.5} & 0.264 & 0.021 & 0.266 & 0.042 & 0.142 & 0.265 \\
    & & & & & 0.895 & -- & -- & -- & -- & 0.105 \\ \hline
    \multirow{12}{*}{0.9} & \multirow{12}{*}{0.5} & \multirow{12}{*}{0.56} & \multirow{2}{*}{1.10} & \multirow{2}{*}{60.0} & 0.567 & 0.010 & 0.161 & 0.020 & 0.169 & 0.073 \\
    & & & & & 0.931 & -- & -- & -- & -- & 0.069 \\
    & & & \multirow{2}{*}{1.15} & \multirow{2}{*}{52.5} & 0.413 & 0.015 & 0.235 & 0.023 & 0.244 & 0.070 \\
    & & & & & 0.875 & -- & -- & -- & -- & 0.125 \\
    & & & \multirow{2}{*}{1.15} & \multirow{2}{*}{60.0} & 0.338 & 0.014 & 0.226 & 0.028 & 0.287 & 0.107 \\
    & & & & & 0.817 & -- & -- & -- & -- & 0.183 \\
    & & & \multirow{2}{*}{1.20} & \multirow{2}{*}{45.0} & 0.425 & 0.019 & 0.208 & 0.022 & 0.245 & 0.081 \\
    & & & & & 0.852 & -- & -- & -- & -- & 0.148 \\
    & & & \multirow{2}{*}{1.20} & \multirow{2}{*}{52.5} & 0.301 & 0.023 & 0.247 & 0.029 & 0.290 & 0.110 \\
    & & & & & 0.776 & -- & -- & -- & -- & 0.224 \\
    & & & \multirow{2}{*}{1.20} & \multirow{2}{*}{60.0} & 0.244 & 0.015 & 0.265 & 0.026 & 0.341 & 0.109 \\
    & & & & & 0.724 & -- & -- & -- & -- & 0.276 \\
    \enddata
    \tablecomments{The first five columns are the properties of the initial conditions. For each of these, two sets of dynamical evolutions are performed; the upper ones is with other bodies of the solar system present, while the lower one has only the resulting bodies and the Sun. The columns $f_\mathrm{Tar}$, $f_\mathrm{Mercury}$, $f_\mathrm{Venus}$ and $f_\mathrm{Mars}$ indicate the fraction of realizations where the runner collides with the target, Mercury, Venus and Mars respectively. $f_\mathrm{Other}$ is the fraction of cases where another collision occur, i.e., either between the largest remnant and another planet, or between two other planets. $f_\mathrm{Rem}$ is the fraction where no collision occur within \SI{20}{\myr}.}
\end{deluxetable*}

For HRC simulations presented in Table~\ref{tab:coll} we perform dynamical evolution of the main remnants. Our interest is in the destination of the runner, i.e. whether a further collision occurs, and if this is the case, whether a return collision, or into a different body, and the relevant impact conditions. We are also interested if there is a relationship between the impact parameters of the successive collisions, including their relative orientations. To provide a realistic environment of late stage of planetary formation, we include other Solar System bodies explicitly in the simulation: Mercury, Venus, Mars, Jupiter, and Saturn, with their present-day characteristics. This can be regarded as representative of past planetary characteristics, although we have not yet studied other configurations. For comparison, we also compute the identical dynamical evolution, but without these additional bodies present. The resulting occurrences are provided in Table~\ref{tab:return-frac}.

\begin{figure}
\centering
\includegraphics{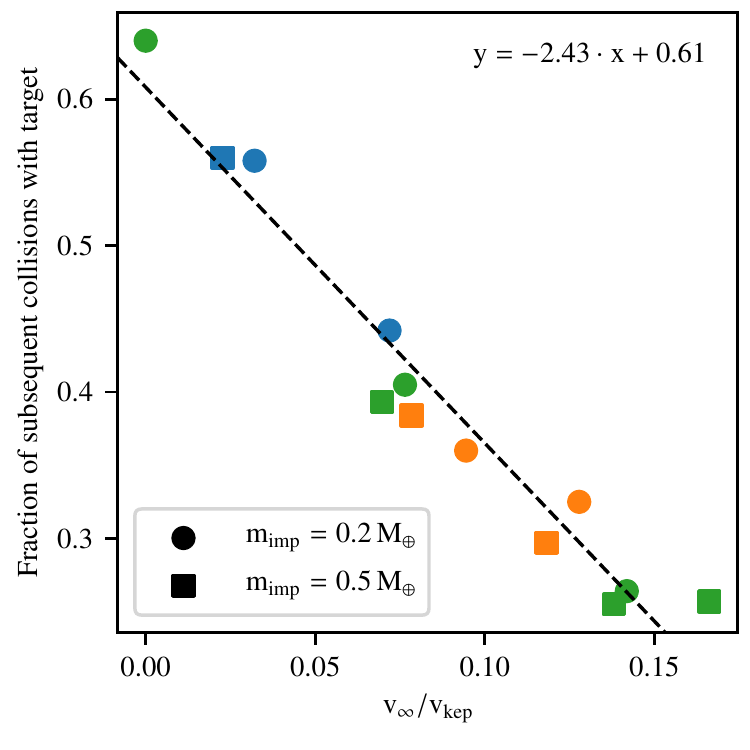}
\caption{Fraction of runners that make a subsequent collision with the target within 20 million years for each set of dynamical evolution calculations, as function of the relative velocity at infinity. The shape of each symbol denotes impactor's mass, with a circle for $m_\mathrm{imp}=\SI{0.2}{\mearth}$ and square for $m_\mathrm{imp}=\SI{0.5}{\mearth}$, and the color for the velocity of the initial collision, with blue for $v_\mathrm{coll}/v_\mathrm{esc}=1.10$, orange for $v_\mathrm{coll}/v_\mathrm{esc}=1.15$ and green for $v_\mathrm{coll}/v_\mathrm{esc}=1.20$. The dashed black line provides the least squres fit of all points, whose parameter are provided on the top right corner. Of the runners that barely escape the collision, only two-thirds return to the target. For characteristic encounter velocities, between 1/3 and 2/3 of them return, decreasing with $v_\mathrm{coll}/v_\mathrm{esc}$.}
\label{fig:return-ftar-vinf}
\end{figure}

With the more realistic case of the addition of planets, the fraction of runners that make a second collision with the largest remnant from the initial collision (the original target) is decreased sharply by nearly a factor of two. As seen in the figure, the value is actually quite dependent on the departing velocity; the lowest velocity runners ($m_\mathrm{imp}=\SI{0.2}{\mearth}$, $v_\mathrm{coll}/v_\mathrm{esc}=1.10$, $\theta_\mathrm{coll}=\SI{55}{\degree}$ with $v_\mathrm{dep}/v_\mathrm{esc}=1.01$ and $m_\mathrm{imp}=\SI{0.2}{\mearth}$, $v_\mathrm{coll}/v_\mathrm{esc}=1.20$, $\theta_\mathrm{coll}=\SI{42.5}{\degree}$ with $v_\mathrm{dep}/v_\mathrm{esc}=1.00$) have the highest return likelihood, with 61.0\% and 60.7\% respectively. On the other hand, the fastest escaping runners have the lowest return probability; in the case of $m_\mathrm{imp}=\SI{0.2}{\mearth}$, $v_\mathrm{coll}/v_\mathrm{esc}=1.20$, $\theta_\mathrm{coll}=\SI{52.5}{\degree}$ with $v_\mathrm{dep}/v_\mathrm{esc}=1.10$, only 27.6\% collide again the same body within \SI{20}{\myr}. In this situation, we actually see a greater likelihood of collision with Venus, at 29.6\%, indicating that for certain accretion regimes, Earth and Venus could have shared some common aspects of their giant impact evolution. And quite noticeably, the intersection at $v_\mathrm{inf}=0$ is lower than unity, at about 0.7, so that the argument that runners mostly return to the same body is  not correct, even just beyond the transition from GMC to HRC. 

We provide in Fig.~\ref{fig:return-ftar-vinf} the fraction of returns versus the relative velocity at infinity, as determined by the two-body problem with $v_\mathrm{inf}^2+v_\mathrm{esc}^2=v_\mathrm{dep}^2$. The principal dependency seems to be on the remaining velocity, although we do note slight shifts for different impactor masses, and different orbital configurations. It should be noted that the escape velocities are not identical between the different points, as the value employed here is the one of the remaining bodies after the collision, which varies due to differences in mass transfer that occur during the encounter. So $v_{\rm esc}$ is always normalized, meaning also that these results can with some degree of confidence be applied to larger and smaller mass regimes, depending on the sensitivity of the problem to the equation of state.

Evidently, just beyond the transition from GMC to HRC, there is a fraction (around 1/3) of lost runners, that become detached from the target. This detachment is more efficient in the presence of other planets. To demonstrate this, we performed the same suite of dynamical evolution studies, but using the longest-period GMC collision, $m_\mathrm{imp}=\SI{0.2}{\mearth}$, $v_\mathrm{coll}/v_\mathrm{esc}=1.10$, $\theta_\mathrm{coll}=\SI{52.5}{\degree}$ (see Table~\ref{tab:transient}). In roughly 75\% of the cases, the secondary body returns after the time obtained from the two-body problem. The remainder take longer to return, with up to 10 years, that is more than two orders of magnitude more than the expected value. This means that with some orbital configuration, the trajectory of the runner is affected by the presence of other planets, and the GMC is close to becoming an HRC. 

In the case of an impact onto the Earth, the transition region between GMC and HRC is narrow. There is only a small window where the impactor is captured, but on a sufficiently wide orbit to be affected by the other bodies. From the two simulations performed for the $m_\mathrm{imp}=\SI{0.2}{\mearth}$ and $v_\mathrm{coll}/v_\mathrm{esc}=1.10$ that are \SI{2.5}{\degree} apart, one is a GMC with the secondary body always returning, and the other in a HRC with $v_\mathrm{dep}/v_\mathrm{esc}=1.01$; so in the case of an Earth-like orbit this window is abrupt, on the order of \SI{1}{\degree}.

The other possibilities behave in the same inverse trend. The decrease of the accretion rate of the runner with increasing departing velocity is a reminiscent of the growth in the oligarchic regime, where the rate decreases as the eccentricities and inclinations of the smaller bodies increase \citep{2001IcarusInaba,2006IcarusChambers}. The departing velocity of the runner is a proxy for the eccentricity of its orbit, so the most rapidly departing ones will spend a lower part of their time in a region where the runner is, while being more likely to reach the other bodies in the system.

We also observe collisions between other bodies than the runner or the target. The fraction of systems that undergo such events are indicated in the column $f_\mathrm{Other}$ of Table~\ref{tab:return-frac}. It could be that the runner collides later on, but as we do not have a consistent collision model in the N-body yet, the state of the system is undetermined. To avoid artifacts in this situation, we halt those simulations at the first collision.

We do not observe any collisions with the central star. Ejections occurs in 1--2\% of the cases. Most of the time, Mars is the ejected planet, while the runner suffers from the same fate in 0.5\% of the cases on average. Ejection from the Solar System following one close encounter with a terrestrial planet is difficult, because the velocity kick provided by such an encounter (which relates to the surface velocity of the planet) is much lower than the escape velocity from the system \citep[e.g.][]{2003ThommesIcarus,2004ApJIdaa}. A possible path towards ejecting the runner is to have it encounter Jupiter, which can provide a sufficient kick to eject the body from the system \citep{2018MNRASJackson}. However the velocity resulting from the collision, and the further encounters with other bodies, are not sufficient to make the runner cross the orbit of Jupiter. Collision with the central star are also unlikely for a similar reason.

The results for the evolution with no other bodies than the collision remnants and the central star are somewhat different. Overall, the return rate is quite high, more than 90\% except in specific cases. Thus the assumption that the runner returns to the same body is valid only without other object present to perturb its orbit. Also, the return rate seems to relate more to the initial collision velocity than the departing velocity.

\subsubsection{Relationship with initial conditions}

\begin{figure}
\centering
\includegraphics{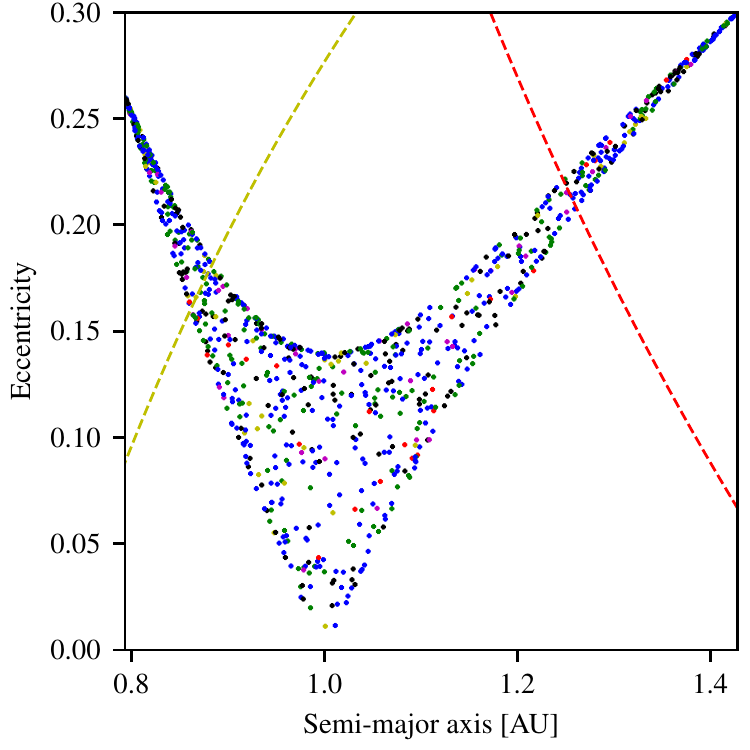}
\caption{Map of outcome with respect to orientation of the initial collision for the case with $m_\mathrm{imp}=\SI{0.2}{\mearth}$, $v_\mathrm{coll}/v_\mathrm{esc}=1.10$, and $\theta_\mathrm{coll}=\SI{60}{\degree}$. The colors are: black for no further collision, blue for a collision between the runner and the target, green for a collision between the runner and Venus, red for a collision between the runner and another body, yellow for a collision between the target and another body than the runner, and magenta for another event. Yellow dashed line denotes Venus-crossing pre-impact orbits, and red dashed line is Mars-crossing pre-impact orbits.}
\label{fig:destination-from-orientation}
\end{figure}

We investigate if specific outcomes are linked to specific initial conditions. For the case with $m_\mathrm{imp}=\SI{0.2}{\mearth}$, $v_\mathrm{coll}/v_\mathrm{esc}=1.10$, and $\theta_\mathrm{coll}=\SI{60}{\degree}$, we plot the type of event obtained with respect to the semi-major and eccentricity map in Fig.~\ref{fig:destination-from-orientation}. There is no obvious correlation between specific pre-impact orbits and type of event obtained after the collision. The same applies to the other sets of dynamical evolution, as in the case where no other bodies are present.

We do not find any restriction on the pre-impact orbits of runner that further collide with other bodies. It is then possible for a body coming from the outer part of the system to produce a HRC and end up colliding further in the inner part of the system, as well as a body to be sent back to its originating part. The same also applies for bodies coming from the inner part of the system. HRC are then able to redistribute material in the system in the same way close encounter are capable.

\subsubsection{Time until subsequent collision}

\begin{figure*}
\centering
\includegraphics{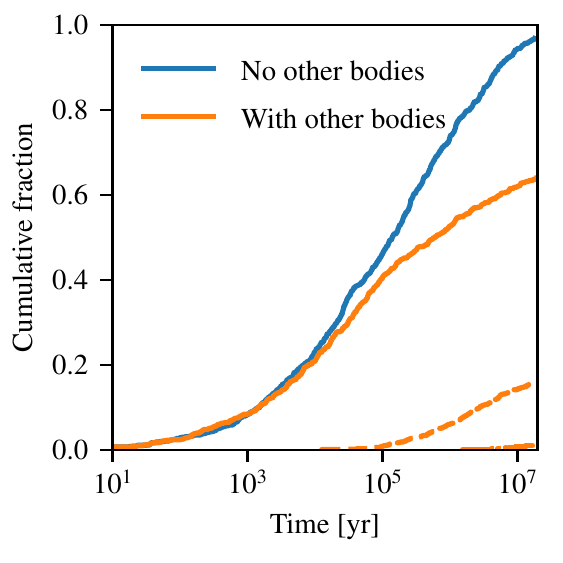}
\includegraphics{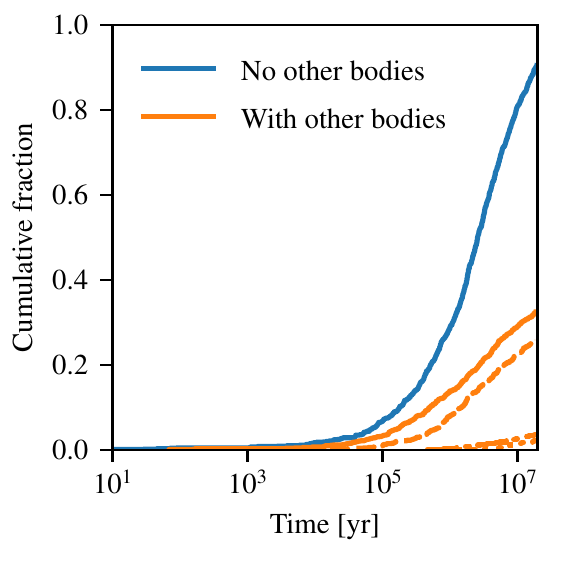}
\includegraphics{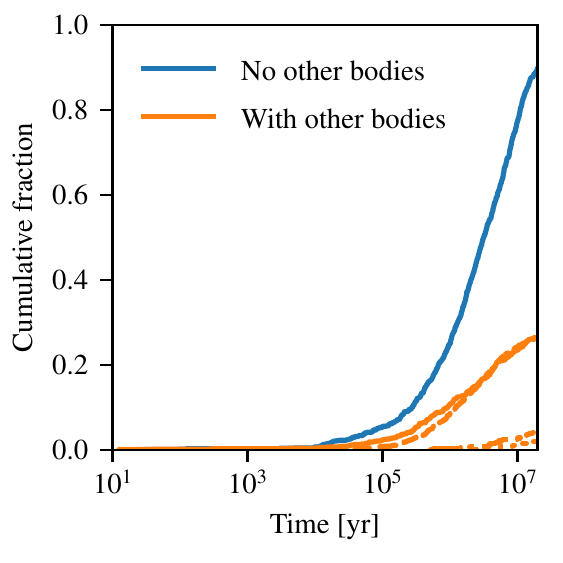}
\caption{A few examples of cumulative distributions of return time and target bodies: $m_\mathrm{imp}=\SI{0.2}{\mearth}$, $v_\mathrm{coll}/v_\mathrm{esc}=1.20$, and $\theta_\mathrm{coll}=\SI{42.5}{\degree}$ (\textit{left}), $m_\mathrm{imp}=\SI{0.2}{\mearth}$, $v_\mathrm{coll}/v_\mathrm{esc}=1.15$, and $\theta_\mathrm{coll}=\SI{60.0}{\degree}$ (\textit{center}), and $m_\mathrm{imp}=\SI{0.2}{\mearth}$, $v_\mathrm{coll}/v_\mathrm{esc}=1.20$, and $\theta_\mathrm{coll}=\SI{52.5}{\degree}$ (\textit{right}). The blue curve is for the dynamical evolution performed without other bodies present while the orange curves are for the ones with Mercury (dash-dotted), Venus (dashed), Mars (dotted), Jupiter, and Saturn (not shown) present. Collisions between the runner and the target are shown with solid curves, while the ones between the runner and other bodies are done according to the previous description.}
\label{fig:return-time}
\end{figure*}

\begin{figure}
\centering
\includegraphics{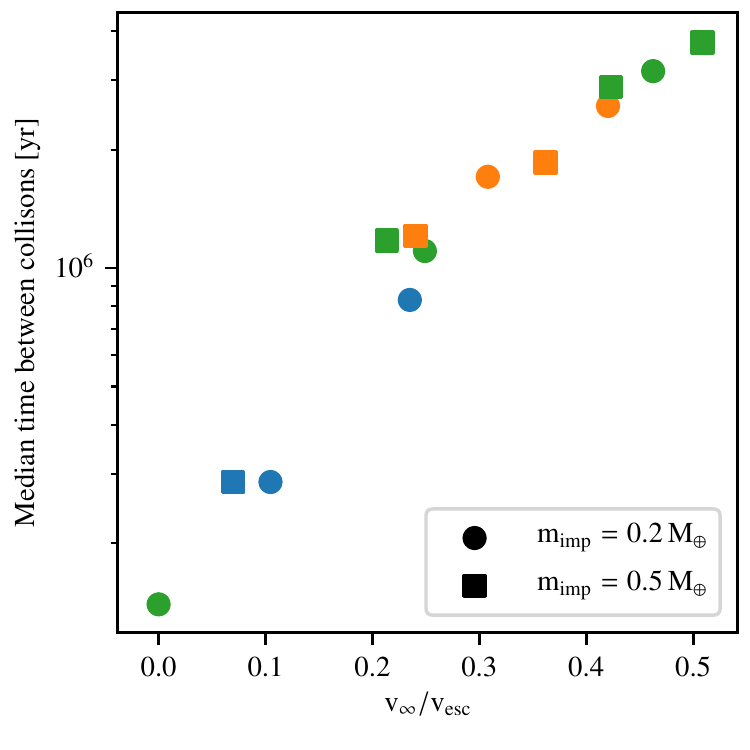}
\caption{Median time between successive collisions for each set of dynamical evolution calculation, as function of the relative velocity at infinity for all dynamical evolution models without other bodies present. Symbols are the same as in Fig.~\ref{fig:return-ftar-vinf}: the shape denotes impactor's mass, with a circle for $m_\mathrm{imp}=\SI{0.2}{\mearth}$ and square for $m_\mathrm{imp}=\SI{0.5}{\mearth}$, and the color for the velocity of the initial collision, with blue for $v_\mathrm{coll}/v_\mathrm{esc}=1.10$, orange for $v_\mathrm{coll}/v_\mathrm{esc}=1.15$ and green for $v_\mathrm{coll}/v_\mathrm{esc}=1.20$.}
\label{fig:return-time-vinf}
\end{figure}

We show the cumulative distribution of the time span between the two collision in Fig.~\ref{fig:return-time} for three series of dynamical evolution. In each series, the blue line shows the outcome when the simulation is restricted to the remnants bodies from the collision and the central star, while the orange curves denote the general case, with other solar system like planets. All the curves are normalized by the initial number of dynamical evolution simulations; so the end value of each curve represent the overall likelihood of the case that is provided in Table~\ref{tab:return-frac}.

As for the return occurrence rate, we obtain the timing is also depending on the departing velocity; runners that are the slowest leaving return fastest. Some runners return as early as a few tens of years, but these are particular cases. Most of them return in a time frame of hundred thousands to a million of years. The correlation between the departing velocity and timing is further consistent with the previous discussion; as the departing velocity increases, so does the runner orbit's eccentricity, and with a higher eccentricity the collision probability decreases. Hence it takes more time.

The presence of the other bodies barely affect the return timing; their major effect remains the reduction on the overall return rate. Further, we note that they affect the later returns; indeed, the cumulative distributions for the return time match pretty well (except in one case), until about \num{e5} years. This figure does not depend on specific collision parameters or outcomes. The perturbations by the other bodies made on the runner, will shift its trajectory over time. The early collisions happen with orbits that are barely disturbed by the other bodies, hence similar collisions are observed in either case.

Collisions with the other bodies do not happen as early as with the target. It takes a similar a similar amount of time than for the perturbations to reduce the rate of collisions with the target, around \num{e5} years, before the first collisions with Venus occur.

\subsection{Return collision parameters}

We further analyze the return collisions properties, which are generally the velocity and angle. In our case we will extend the analysis to the relationship between the successive collisions and also the orbital configuration. This ultimately enables us to determine their regime, i.e. whether these are HRC again or mergers.

For reference, we modelled -- by mistake -- a series of dynamical evolution where all the initial collisions were assumed to happen vertically, i.e. $\theta\approx0$. In case of subsequent events and timing, it behaves as for coplanar impacts; however it will be useful further on, as it provides insight on the effect of geometry on the results.

\subsubsection{Velocity}
\label{sec:ret-vel}

\begin{figure*}
\includegraphics{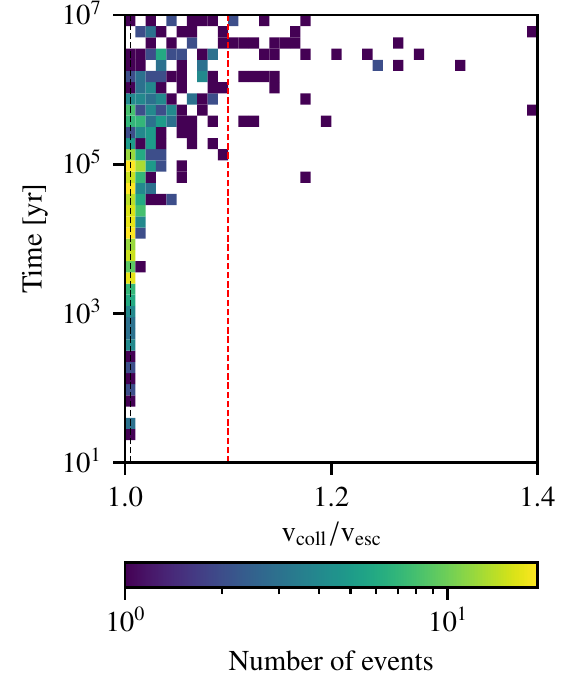}
\includegraphics{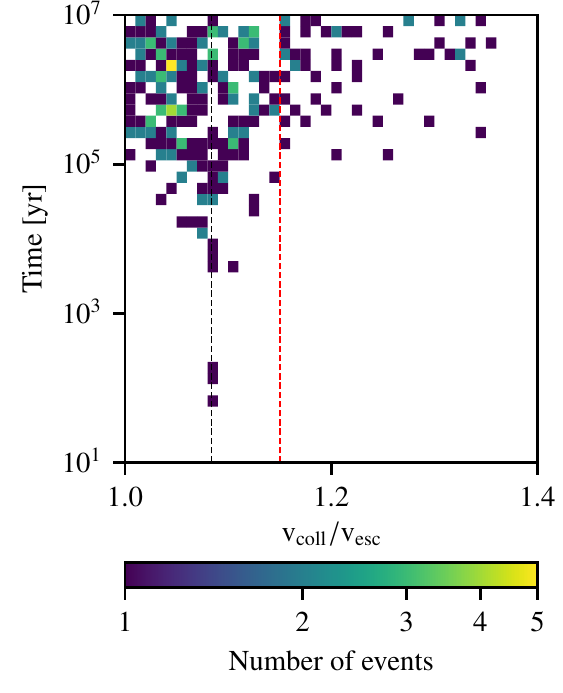}
\includegraphics{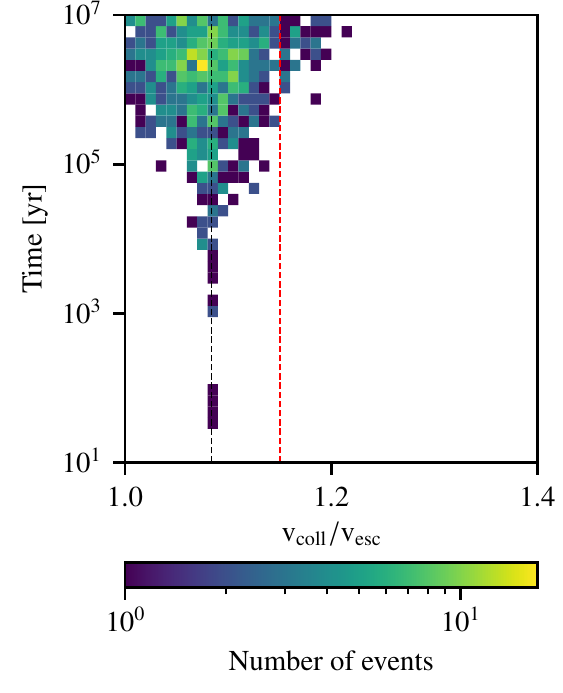}
\caption{2D histogram of impact velocity versus time span between successive impacts for return collision on the same body. \textit{Left}: $m_\mathrm{imp}=\SI{0.2}{\mearth}$, $v_\mathrm{coll}/v_\mathrm{esc}=1.10$ and $\theta_\mathrm{coll}=\SI{55}{\degree}$ with other solar system bodies present; \textit{Center}: $m_\mathrm{imp}=\SI{0.2}{\mearth}$, $v_\mathrm{coll}/v_\mathrm{esc}=1.15$ and $\theta_\mathrm{coll}=\SI{60}{\degree}$ also with other solar system bodies present; \textit{Right}: same, but only with the collision remnants and the central star. The initial impact velocity is shown with a dashed red line while the impact departure velocity is shown with a dashed black line.}
\label{fig:return-velocity}
\end{figure*}

Velocity of collision is a key parameter determining the outcome, e.g., whether it is a HRC, or a disruption, or a GMC, or a merger. Thus the velocity of the returning collision is of key interest in determining the final evolution of a giant impact chain, whether the chain continues (another HRC) or ends (a merger). In this section we correlate the incoming velocity for the first collision, with the return velocity of the second collision. For example, for immediate returns the return velocity is expected to be close to the departing velocity. But when the return collision comes later, mutual interactions might affect the trajectory leading to different impact velocities. Therefore the results are shown as 2D histograms of impact velocity versus delay until the subsequent collision occurs.

We show such results for three sets of dynamical evolution in Fig.~\ref{fig:return-velocity}. The impact and departure velocities of the initial collision are marked with red and black dashed lines respectively. We do observe a few features: the return velocity is more related to the departing velocity than the initial velocity, and as the time between the collisions increase, so does the spread in impact velocities. The cases shown here represent the variety of outcomes obtained in this study. The first case is where the departing velocity is close to the mutual escape velocity, i.e. the collision is close to the transition between HRC and GMC. Only the case with the other solar system bodies is shown, as without them, all return event occur at the same velocity. The second and third cases show a departing velocity quite higher than the mutual escape velocity where spread of the return collision velocity can be obtained in both directions. With other bodies present, we note that there a few events with $v_\mathrm{coll}/v_\mathrm{esc}>1.25$, which is not the case otherwise. Nevertheless, as the departing velocity is lower than the initial collision velocity, when see that there is only a small fraction of the return collision occurring at an higher velocity than the initial one. The velocity decrease is less pronounced as the impact angle of the initial collision increase; the collision dissipates less energy for grazing events so the departing velocity is closer to the initial impact velocity, and so does the impact velocity of the return collision.

The relationship between the departing and return velocity is quite straightforward. If energy was conserved in the point of view of two bodies (target and runner), then the return collision velocity would happen with same velocity. With other bodies, close encounters lead to modifications of the orbits, and the return velocity can vary. The presence of other bodies provides more means of exchanges, and therefore the impact velocity spread in higher in this situation.

Second, the higher spread over time is linked to the same effect. As time goes, perturbations by the other bodies increase. The early returns happen at a velocity very close to the departing velocity. After again roughly the same time as given in the previous section, \num{e5} years, the impact velocities start to diverge noticeably. In the case with other bodies present, the median of the distribution remains at the departing velocity.

To properly resolve the return collision, correct determination of the departing velocity from the initial collision is required, as this is what determines the properties of the second event. Assuming that the runner leaves the target with essentially the same velocity as the one it came with neglects various physical processes, among them energy dissipation through shocks.

\begin{figure}
\centering
\includegraphics{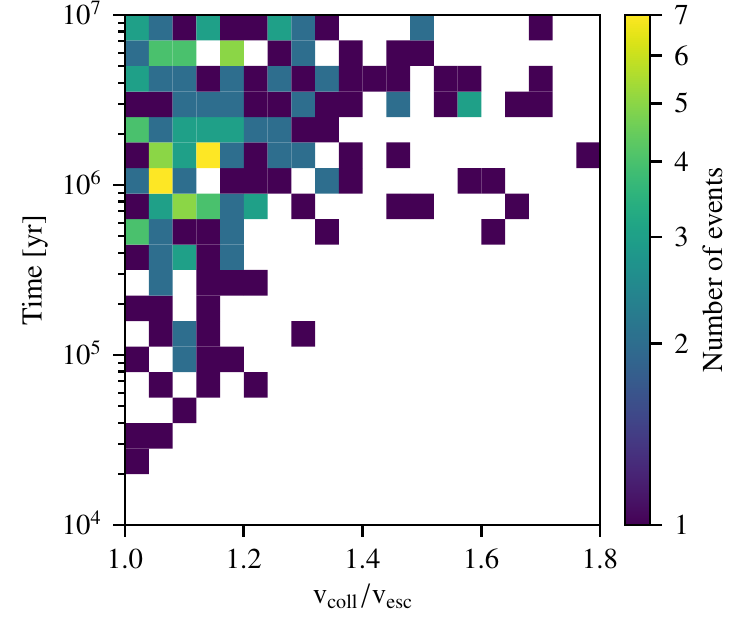}
\caption{2D histogram of impact velocity versus time span between a first collision hitting the target and a second collision hitting Venus. $m_\mathrm{imp}=\SI{0.2}{\mearth}$, $v_\mathrm{coll}/v_\mathrm{esc}=1.15$ and $\theta_\mathrm{coll}=\SI{60}{\degree}$. The impact velocity, assuming the same velocity at large separation as when departing from the initial collision, is shown with a dashed black line.}
\label{fig:venus-velocity}
\end{figure}

Since in the cases with a relatively large departing velocity many collisions between the runner and Venus are obtained, we perform the same analysis for that series of events. The result is shown in Fig.~\ref{fig:venus-velocity}. It should be noted that the axis are not showing the same values as in Fig.~\ref{fig:return-velocity}; the time axis has been shrunk to show only the later times, as no early collisions occur and the velocity axis has been expanded, and there only an handful of event that happen outside of the range shown here. We observe a similar trend than the return collision velocity increases over time. However, the base velocity does not seem to be related to the departing velocity of the prior collision; rather it seems to be somewhat lower at around $v_\mathrm{coll}/v_\mathrm{esc}\approx1.05$. On Fig.~\ref{fig:venus-velocity}, the velocity has been corrected to account for the different mass and radius of the Venus body compared to the end state of the target.

If we follow the methodology from \citet{2018MNRASJackson} again, the minimum relative velocity at Venus of an impactor that has an Earth-crossing orbit (assuming both Venus and Earth are on circular orbits) results in an impact velocity of $v_\mathrm{coll}/v_\mathrm{esc}\simeq1.045$. This value is very close to the one reported above, so the results are consistent. The same orbit requires a relative velocity at Earth's orbit of $v_\mathrm{rel}/v_\mathrm{esc}\simeq0.27$, or $v_\mathrm{dep}/v_\mathrm{esc}\simeq1.04$. This implies that runner departing with a lower velocity cannot have directly Venus-crossing orbits, and require interactions with other bodies to become so. As the departing velocity increase, so does the fraction of orbits that can be Venus-crossing (Figure~\ref{fig:orb}).

We also obtained collision with the Mercury and Mars analogues, however the number of those events is too low to obtain a meaningful statistics.

\subsubsection{Impact angle}

\begin{figure*}
\centering
\includegraphics{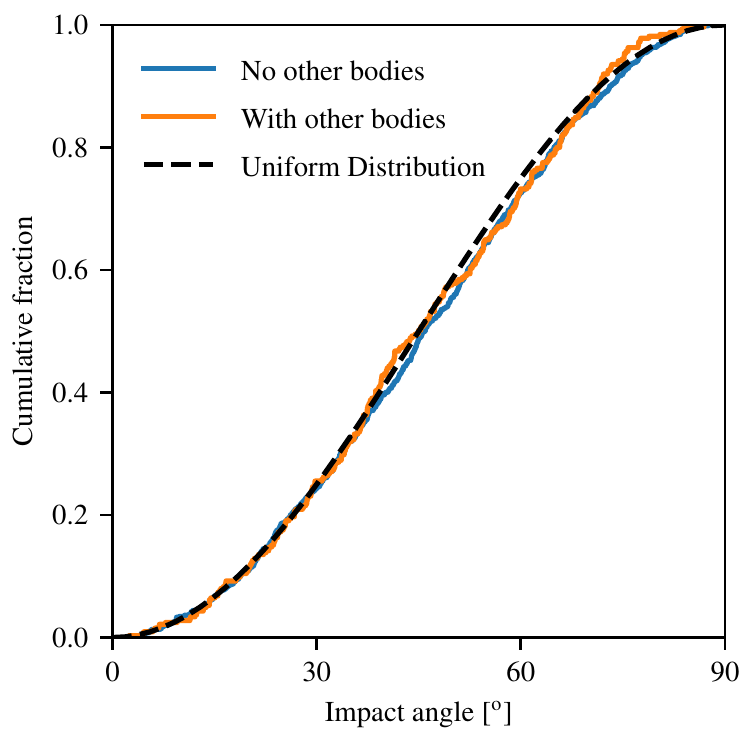}
\includegraphics{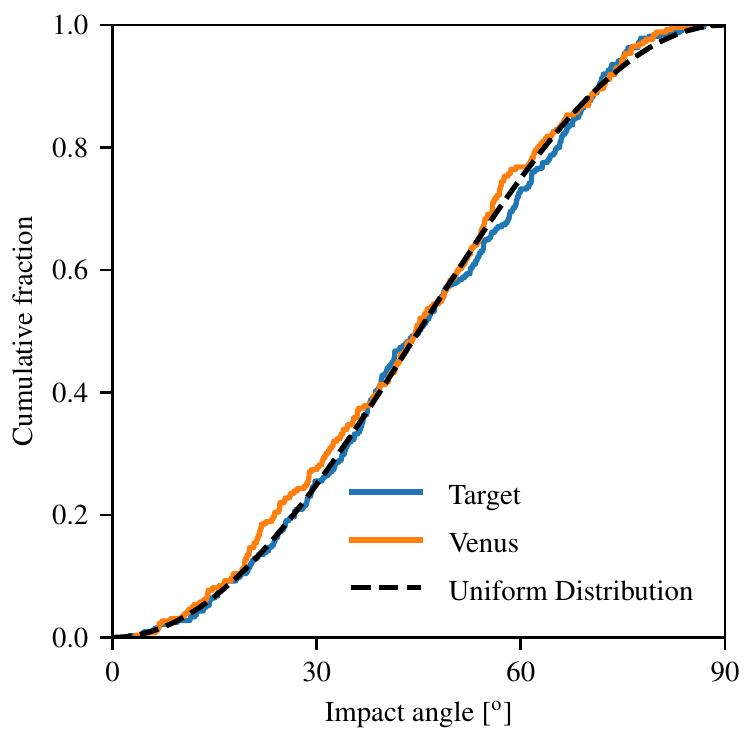}
\caption{Cumulative distribution of impact angles for collision between the runner and target (\textit{left}) and the runner and Venus (\textit{right}). Results from the dynamical evolution following the collision with $m_\mathrm{imp}=\SI{0.2}{\mearth}$, $v_\mathrm{coll}/v_\mathrm{esc}=1.15$ and $\theta_\mathrm{coll}=\SI{60}{\degree}$. The black dashed line is the expected distribution if the offset of the runner follows a uniform distribution, computed following \citet{1962BookShoemaker}.}
\label{fig:impact-angle}
\end{figure*}

The basic collision geometry is set by the impact angle. Its cumulative distribution for the return collisions with the target in one dynamical evolution series, both in the cases with and without additional bodies is shown in Fig.~\ref{fig:impact-angle}. Along with, we provide a theoretical distribution of impact angles if impactors were uniformly approaching the target following \citet{1962BookShoemaker}. In the latter case, the mean, median and mode of the distribution are all located at \SI{45}{\degree}.

For all the dynamical evolution models performed, with additional bodies present or not, we obtain very similar results: the distribution of impact angles of the return collision is compatible with the uniform distribution. To quantify this, we run a one-sample Kolmogorov--Smirnov test (KS) to asses the compatibility of the results with the reference distribution. All but one obtained \textit{p-values} are greater than 3\%, so we have no strong argument to reject that the obtained distribution derive from a uniform distribution. The exception is the case with $m_\mathrm{imp}=\SI{0.2}{\mearth}$, $v_\mathrm{coll}/v_\mathrm{esc}=1.20$ and $\theta_\mathrm{coll}=\SI{52.5}{\degree}$, which shows an excess of collisions with $\theta\approx\SI{30}{\degree}$.

The compatibility with a uniform distribution implies that the end geometry of the prior collision has no influence whatsoever of the subsequent event, and in this direction, the fact that the same bodies had already collided has no importance. Since the target-runner pair is not a isolated system, there is no reason to expect that angular momentum would be conserved, which would be needed in order to set a constraint on the impact angle of the return collision. Since HRC are happening for grazing angles, the return collision has on average a steeper impact angle.

The same compatibility with uniform distribution is also obtained for collisions with other bodies. This is not much of a surprise, as if the distribution of impact angles for a return collision with the same body is compatible with a uniform distribution, it would be hard to understand why this would not be the case for other bodies.

Lastly, there is a mismatch for the set of dynamical evolution where all the collision were modelled as occurring in the orbital plane. If all collisions happen in the same plane, then a different distribution of impact angles, shifted towards steeper values, is expected \citep{2017PhDTEmsenhuber,2018NatAsLeleu}. In the latter case, more head-on collisions are favored, and the median is located at \SI{30}{\degree}, compared to \SI{45}{\degree} in the general case following \citet{1962BookShoemaker}. In our case we observe a slight increase of more head-on events, though the distribution of impact angles is closer to the general distribution than the one for impacts in one plane. As we shall discuss further on, the other properties of this dynamical evolution set tend to the expected distributions for impact uniformly oriented in space. As to why, we can say that our SPH simulations are not exactly aligned in the direction orthogonal to the impact plane; the relative motion of the two remnants is inclined by nearly \SI{1}{\degree}. This small deviation is sufficient to induce vertical motion of the bodies that results in a uniform distribution of impact angles in a volume rather than in a plane. Only a small vertical deviation in the vertical direction is needed to obtain the general distribution, slightly greater than $r_\bot$. We calculated that the required inclination to fulfill this condition is on the order of \num{e-3}. Since the return velocity is somewhat smaller than in the initial collision, $r_\bot$ is increased; however the value remains small enough to have a uniform-in-space-like distribution.

\subsubsection{Alignment between successive collisions}

\begin{figure}
\centering
\includegraphics{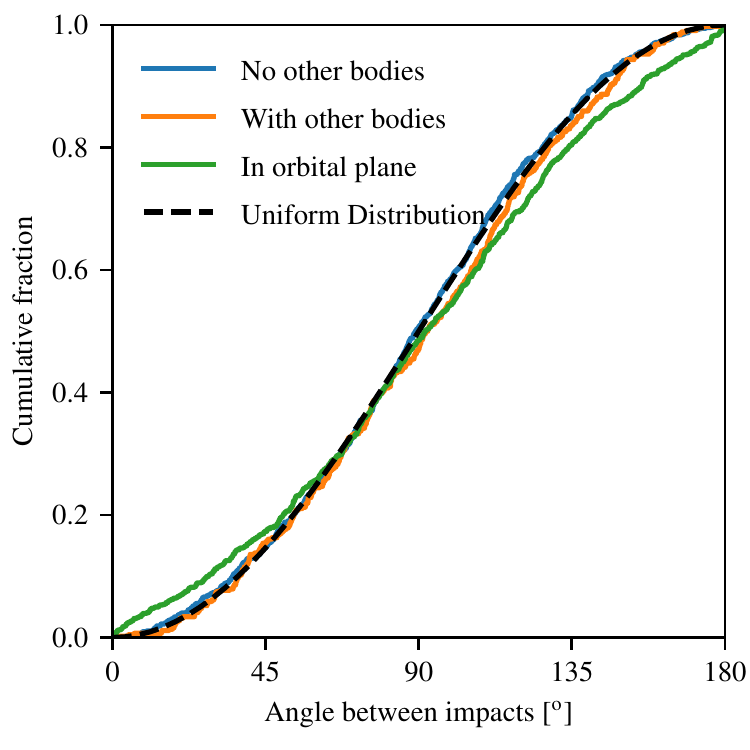}
\caption{Cumulative distribution of angles for collision between the runner and target. Results from the dynamical evolution following the collision with $m_\mathrm{imp}=\SI{0.2}{\mearth}$, $v_\mathrm{coll}/v_\mathrm{esc}=1.10$ and $\theta_\mathrm{coll}=\SI{60}{\degree}$. The black dashed line is the expected distribution if the offset of the runner follows a uniform distribution.}
\label{fig:impact-alignment}
\end{figure}

In the cases the runner returns to the target, there is a series of two collisions between the same bodies. It is then possible to search for relationships between these events, such as the spatial alignment of the impact planes. HRC being grazing events, a part of the relative angular momentum is transferred during the collision from the orbital motion into spin. If we assume the spin axis is not modified in between the collision, then the alignment between the relative angular momentum vectors is a proxy for the alignment of the returning runner and the spin of the bodies. The results for one series of dynamical evolution is provided in Fig.~\ref{fig:impact-alignment}; the other series behave similarly.

As for the impact angle, we find that the angle between the specific angular momentum vectors is compatible with a uniform distribution, except for the cases where the initial collision is assumed to happen in the orbital plane. The lowest \textit{p-value} obtained (with the exception of the in orbital plane configuration) is 2\%, with a even distribution of values up to unity.

The case where all collisions are assumed to be in the orbital plane is special, as we observe an excess of collisions occurring with a relative angle close to either \SI{0}{\degree} or \SI{180}{\degree}; i.e. also in the same plane, with prograde and retrograde alignment respectively. This is again consistent with the previous result about impact angle, as the collisions with relative angle happen close to the orbital plane. This seems to be restricted to the specific case of collisions in the orbital plane however. In our other set where the are only almost-vertical impacts, we do not find such a correlation in the angles between successive collisions; the distribution is similar to the ones from the general case and compatible with a uniform distribution. Coplanar collisions are special, in the sense that angular momentum is almost perpendicular to the orbital plane, therefore vertical motion is limited. It has a negligible effect on the overall distribution, as the probability of obtaining such a case -- or a close-by one -- is infinitesimal: the probability of a certain inclination is $\mathrm{d}P\propto\cos{i}\mathrm{d}i$ assuming a uniform distribution of the specific angular momentum, and $\cos{(-\pi/2)}=\cos{(\pi/2)}=0$. Therefore, we do not find any evidence for correlation between the alignment of successive collisions between the same bodies, except in specific circumstances which require some external mechanism, e.g. inclination damping by a gas disk, to be attained.

If the orientation of the previous collision relates to the spin axis of the bodies prior the return collision, then no specific configuration is favored. Most of the event should occur with the impactor colliding perpendicular to the spin motion of the target. Whether the impact occurs at the pole or on the equator is unconstrained as well. Note that we usually refer to the spin motion of the target only, though the impactor may be rotating as well, since the target with its larger mass carries an important fraction of the spin angular momentum. Prograde and retrograde collisions have equal likelihood, but it is unlikely that these are well aligned.

The non-alignment makes HRR quite different from GMC. In GMC, the runner returns after a short period of time with same alignment and angle than the end state of the initial collision. Neither holds in HRR, so that the overall accretion process behaves differently. Both HRC and GMC carry significant angular momentum; usually higher than the spin angular momentum a body can sustain \citep{1969BookChandrasekhar}. In the case of GMC, a part of the material must be ejected, carrying away angular momentum \citep[e.g.][]{2013IcarusAsphaug}. For HRR, there are other ways to limit the final spin angular momentum. First the modification of the impact angle; we saw in the previous section that the impact angle of the return collision is unconstrained by the end state on the initial collision. Since the equivalent of the impact angle is quite large and the one of the return collision is on average $\SI{45}{\degree}$, this leads to a reduction of the angular momentum carried by the orbital configuration. Second, if the return collision occurs perpendicularly to spin axis of the body, then angular momentum is added in quadrature, which leaves a final value lower than if the return collision happen in the prograde direction, which is the case of a GMC. HRR can then be more efficient to accrete the returning runner, provided the return collision is merger, either simple or GMC.

\subsubsection{Orientation}

\begin{figure*}
\centering
\includegraphics{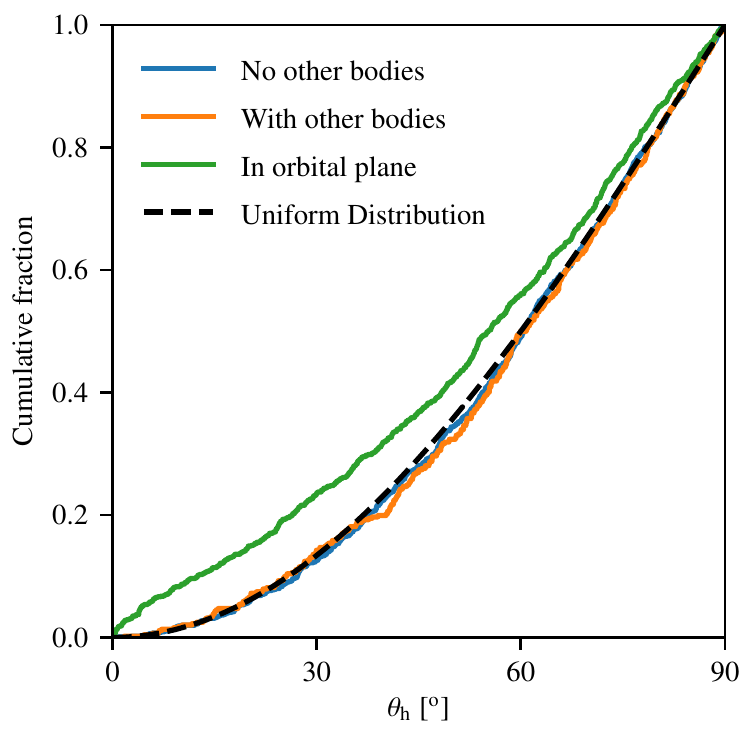}
\includegraphics{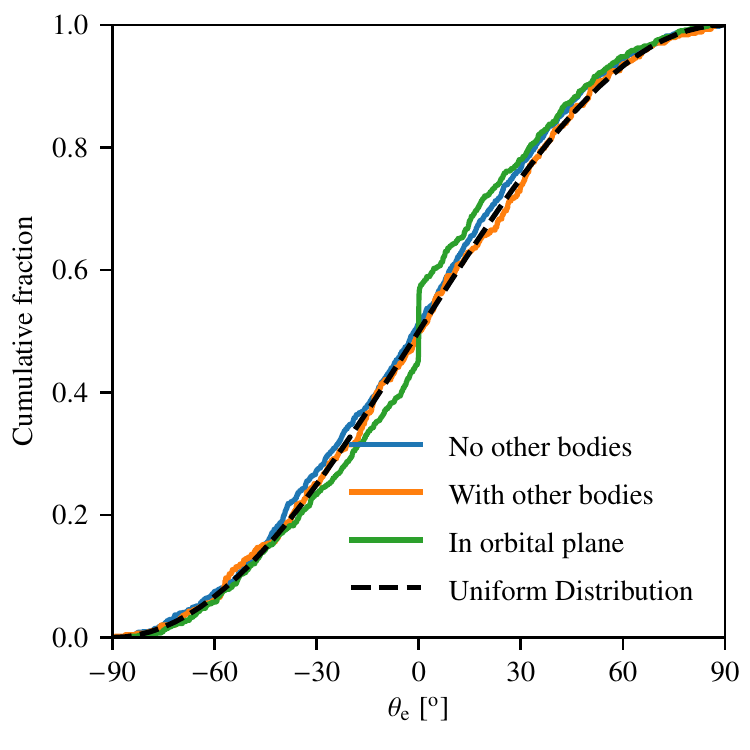}
\caption{Cumulative distribution of angles for the inclination of the impact plane (\textit{left}) and the relative velocity vector at infinity (\textit{right}). Results from the dynamical evolution following the collision with $m_\mathrm{imp}=\SI{0.2}{\mearth}$, $v_\mathrm{coll}/v_\mathrm{esc}=1.10$ and $\theta_\mathrm{coll}=\SI{60}{\degree}$. The black dashed line is the expected distribution if the offset of the runner follows a uniform distribution.}
\label{fig:eccvect-inclination}
\end{figure*}

\begin{figure}
\centering
\includegraphics{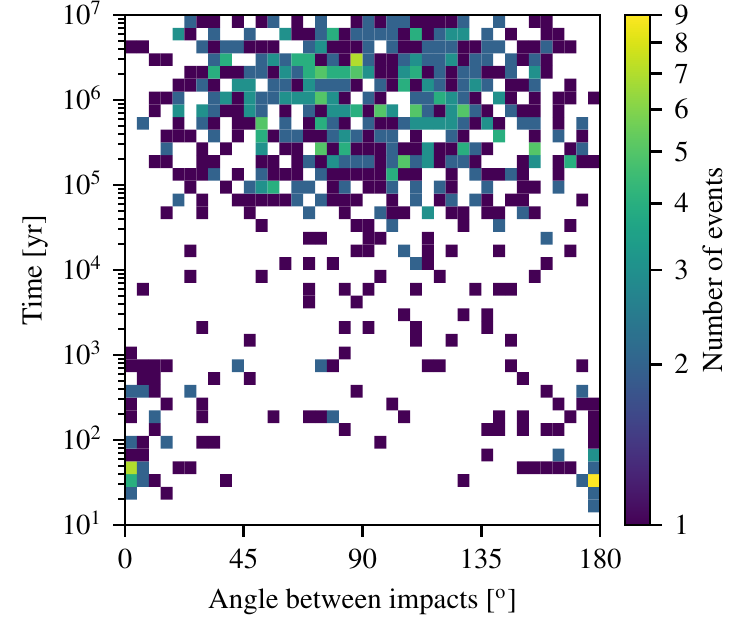}
\caption{2D histogram of the angle between successive collision and time for the case with $m_\mathrm{imp}=\SI{0.2}{\mearth}$, $v_\mathrm{coll}/v_\mathrm{esc}=1.10$ and $\theta_\mathrm{coll}=\SI{60}{\degree}$ and all initial collision assumed to occur in the orbital plane. At early time, the orientation of the return collision is correlated to the first collision. Later the return collision is what one expects from a random distribution of impacts.}
\label{fig:impact-alignment-correl}
\end{figure}

To determine the orientation of the collision, we check two vectors: specific angular momentum, and relative velocity at infinity (or the eccentricity vector in the unlikely case of a closed orbit). They are orthogonal by construction, though they do not provide the same information. For instance, an impact occurring in the orbital plane is rare, as we saw in the previous sections. However a small vertical offset when compared to orbital distance is able to produce highly oblique impacts (see related discussion in section~\ref{sec:methods-dyn}). The relative velocity vector at infinity on the other hand provides a greater insight on the mutual inclinations, as a non-horizontal one requires some inclination \citep{2018MNRASJackson}.

The results are shown in Fig.~\ref{fig:eccvect-inclination} for the dynamical evolution following the collision with $m_\mathrm{imp}=\SI{0.2}{\mearth}$, $v_\mathrm{coll}/v_\mathrm{esc}=1.10$ and $\theta_\mathrm{coll}=\SI{60}{\degree}$. While not displayed here, other cases have been performed, and show similar results. For the initial conditions that where taken to follow a uniform distribution in space, we do not have sufficient evidence against the results being compatible with the same. We do obtain a p-value of \num{3e-3} for inclination of the relative velocity vector at infinity in one set, but otherwise the values are quite higher. This is not the case when the initial collision is assumed to occur in the orbital plane. The results are not compatible with a uniform, with p-values between $\num{e-6}$ and $\num{e-3}$, both for both quantities. Looking in more details, we note that there is a contribution of roughly 10\% of collisions that happen very close to the orbital plane, as the inclination of the relative velocity vector at infinity is almost zero and the inclination of the collision plane is low. If we plot a 2D histogram of the angle between the collision plane versus time span between collisions, as in Fig.~\ref{fig:impact-alignment-correl}, it can be seen that this contribution is from early returns. These events happen on so short time scales, less than a thousand years, that they could not have sufficient close encounters to modify their orbit so that the mutual inclination required for collisions outside of the impact plane is obtained. Also these early returns are quite evenly distributed between angles close to \SI{0}{\degree} (prograde) and \SI{180}{\degree} (retrograde). On the other hand, later return follow more closely the expected values for a uniform distribution, so the non-compatibility is linked to a small set of early return collisions.

These results permit us to determine the relevance of our initial conditions, that we selected so that the relative velocity vector at infinity is uniformly distributed in space. The same distribution is the natural outcome of dynamical evolution. Even when all the initial collisions happen in the orbital plane, the majority of return collision follow the same uniform distribution, with the exception of about 10\% of event due to the specific geometry. We also observe a uniform distribution in our special set that assumes all initial collisions to happen in a near-vertical direction. Therefore the initial conditions for the dynamical evolution are representative for the expected distribution of impact orientations. Furthermore, we verified that there is no temporal evolution of the distribution. Early collisions are also compatible with a uniform distribution.

\subsection{Effect of initial eccentricity}

The initial conditions for the dynamical evolution assume the target is on a circular orbit prior to the giant impact, and the orbit of the impactor is computed to obtain the desired collision orientation. Since the target is also deflected during a collision, its orbit at the beginning of the dynamical evolution is elliptical. The same applies to the center of mass of the target-impactor system. The trajectory of the center of mass depends principally on the impactor's mass and the relative velocity, and only weakly on the impact angle. The cases with the larger impactor, $m_\mathrm{imp}=\SI{0.5}{\mearth}$, therefore have an overall larger eccentricity of remnants than for the smaller one, $m_\mathrm{imp}=\SI{0.2}{\mearth}$.

We check the orbits of the remnants immediately after the collision, to check if they are initially crossing with other bodies. Our choice of pre-impact orientation results in similar eccentricities for both remnants; the figures we report here are thus valid for both remnants. In the cases with the smaller impactor, $m_\mathrm{imp}=\SI{0.2}{\mearth}$, the orbits are not intially crossing Venus for the slow departing runners, while some are for the faster ones, up to 18\%. We therefore always obtain more collisions with Venus than there are of initial crossing orbits. Mutual excitation are therefore needed to obtain a good part of the collisions with Venus. For the larger impactor, $m_\mathrm{imp}=\SI{0.2}{\mearth}$, the figures of crossing orbit fractions are similar than for the smaller at same velocity and angle, albeit slightly lower. The increased rate is then not due to more eccentric orbits.

To verify the effect of the choice of the initial orbital setup, we perform a second dynamical evolution were it is assumed that the center of mass in on a circular orbit, for the collision with $m_\mathrm{imp}=\SI{0.5}{\mearth}$, $v_\mathrm{coll}/v_\mathrm{esc}=1.20$ and $\theta_\mathrm{coll}=\SI{45}{\degree}$. The main effect is to reduce the overall eccentricity of the remnants, therefore keeping them further away from the other bodies while increasing their potential encounters. We observe two main consequences: the number of collisions between the two remnants in increase from 0.425 to 0.552 and the return collision velocity is reduced so that the relative difference to the departing velocity is about \num{e-4}. The other results are in good agreement with the ones discussed previously.

Assuming that the target is on a circular orbit prior to the collision hence does not lead to a special situation. The deflection occurring during the collision lead to similar eccentricities in both bodies. However, the eccentricities are even lower when the center of mass is assumed to be on a circular orbit, and in this case the return rate increases while lower return collision velocities are obtained. The differences between the two cases remain limited still. The collisions with the larger impactor, $m_\mathrm{imp}=\SI{0.5}{\mearth}$, result in slighter higher eccentricities. This can in part explain the increased rate of collision with other bodies compared to the smaller impactor.

\subsection{Return collision type}

\begin{deluxetable*}{ccccccccccccc}
	\tablecaption{Type of return collisions for subsequent collision obtained in the dynamical evolution with the other solar system bodies present.\label{tab:return-type}}
	\tablehead{
		\colhead{$m_\mathrm{tar}$ [\si{\mearth}]} & \colhead{$m_\mathrm{imp}$ [\si{\mearth}]} & \colhead{$\gamma$} & \colhead{$\frac{v_\mathrm{coll}}{v_\mathrm{esc}}$} & \colhead{$\theta_\mathrm{coll}$ [\si{\degree}]} & \colhead{$f^\mathrm{ini}_\mathrm{HRC}$} & \colhead{$f^\mathrm{SL12}_\mathrm{Merge}$} & \colhead{$f^\mathrm{SL12}_\mathrm{GMC}$} & \colhead{$f^\mathrm{SL12}_\mathrm{HRC}$} & \colhead{$f^\mathrm{SL12}_\mathrm{Dis}$} & \colhead{$f^\mathrm{C}_\mathrm{Merge}$} & \colhead{$f^\mathrm{C}_\mathrm{GMC}$} & \colhead{$f^\mathrm{C}_\mathrm{HRC}$}}
	\startdata
	\multirow{7}{*}{0.9} & \multirow{7}{*}{0.2} & \multirow{7}{*}{0.22} & 1.10 & 55.0 & 0.33 & 0.00 & 0.57 & 0.05 & 0.38 & 0.39 & 0.54 & 0.07 \\
	& & & 1.10 & 60.0 & 0.33 & 0.00 & 0.51 & 0.14 & 0.35 & 0.33 & 0.47 & 0.20 \\
	& & & 1.15 & 52.5 & 0.47 & 0.00 & 0.44 & 0.19 & 0.37 & 0.35 & 0.37 & 0.28 \\
	& & & 1.15 & 60.0 & 0.47 & 0.00 & 0.42 & 0.22 & 0.36 & 0.28 & 0.40 & 0.32 \\
	& & & 1.20 & 42.5 & 0.54 & 0.04 & 0.58 & 0.06 & 0.32 & 0.40 & 0.52 & 0.08 \\
	& & & 1.20 & 45.0 & 0.54 & 0.00 & 0.42 & 0.16 & 0.42 & 0.41 & 0.39 & 0.20 \\
	& & & 1.20 & 52.5 & 0.54 & 0.00 & 0.33 & 0.21 & 0.46 & 0.43 & 0.27 & 0.30 \\ \hline
	\multirow{6}{*}{0.9} & \multirow{6}{*}{0.5} & \multirow{6}{*}{0.56} & 1.10 & 60.0 & 0.25 & 0.00 & 0.65 & 0.06 & 0.29 & 0.18 & 0.65 & 0.17 \\
	& & & 1.15 & 52.5 & 0.43 & 0.00 & 0.57 & 0.10 & 0.33 & 0.17 & 0.56 & 0.27 \\
	& & & 1.15 & 60.0 & 0.43 & 0.00 & 0.53 & 0.15 & 0.32 & 0.15 & 0.52 & 0.33 \\
	& & & 1.20 & 45.0 & 0.53 & 0.00 & 0.59 & 0.14 & 0.27 & 0.15 & 0.54 & 0.31 \\
	& & & 1.20 & 52.5 & 0.53 & 0.00 & 0.48 & 0.23 & 0.29 & 0.15 & 0.41 & 0.44 \\
	& & & 1.20 & 60.0 & 0.53 & 0.00 & 0.48 & 0.27 & 0.25 & 0.08 & 0.44 & 0.48 \\
	\enddata
\tablecomments{The first five columns are the properties of the parent collision. $f^\mathrm{ini}_\mathrm{HRC}$ is an estimate of the fraction initial-like collision being in the HRC regime (assuming the same parameters, except for an impact angle following the \citealp{1962BookShoemaker} distribution). The following four column, $f^\mathrm{SL12}_\mathrm{Merge}$, $f^\mathrm{SL12}_\mathrm{GMC}$, $f^\mathrm{SL12}_\mathrm{HRC}$ and $f^\mathrm{SL12}_\mathrm{Dis}$ are the fraction of subsequent collision between the runner and the target that are in the merging, GMC, HRC and Disruption regimes, as assessed following \citet{2012ApJStewart}. The last three columns, $f^\mathrm{C}_\mathrm{Merge}$, $f^\mathrm{C}_\mathrm{GMC}$, and $f^\mathrm{C}_\mathrm{HRC}$ are the same, but following \citet{2019ApJCambioni}; no collisions were found to be in the disruption regime with this procedure.}
\end{deluxetable*}

We finish the analysis of the HRR by determining the type of collision they would produce. Direct modelling of each event with SPH would bear a prohibitive computational cost, therefore we asses the type of those events using scaling laws. We use both the scaling of \citet{2012ApJStewart}, hereafter \citetalias{2012ApJStewart}, which is based on \citet{2012ApJLeinhardt} with the modification that in the grazing regime, there is a further division between GMC and HRC based on the results from \citet{2010ApJKokubo}, and by \citet{2019ApJCambioni}. The results from this analysis are provided in Table~\ref{tab:return-type}. For comparison, we provide the likelihood for the initial collision to be HRC, $f^\mathrm{ini}_\mathrm{HRC}$, by estimating the impact angle for which transition between GMC and HRC occurs for the given bodies and impact velocity using the results from Table~\ref{tab:coll}, and then using the uniform distribution in space following \citet{1962BookShoemaker} to asses the probability that the impact angle is greater than that value.

The results using both methods are in good agreement when it comes to the HRC and GMC regimes, with some differences for the fraction of HRC in most massive impactor. Quite a number of collisions are found to be close to the transition between GMC and HRC, hence there is some uncertainty associated with the classifcation performed in \citet{2019ApJCambioni}. For the others, the discrepancy between ``merge'' and ``disruption'' is due to the different classification. The disruption regime in \citetalias{2012ApJStewart} includes partial accretion onto the target, while the merging regime in \citet{2019ApJCambioni} also encompasses the part where some material is ejected in the form of debris. These two categories should be seen as collisions that do not produce a second remnant, transient or not.

We obtain that when an HRR occurs, it is more likely to be a merger (direct or GMC) than the previous collision. This relates to the decrease of the relative velocity following the initial HRC. As the HRR impact velocity is lower than in the previous collision, the transition shifts to higher impact angles, therefore reducing the probability of subsequent HRC. Slowly departing runners, close to the mutual escape velocity have a low probability of subsequent HRC ($<10\%$), while for the more rapid ones, this figure increases to roughly 30\%. Series of HRC collisions would then mostly end after a few events. This result is consistent with the outcome from \citet{2013IcarusChambers}, which found that accretion time scales of terrestrial planets are increased by a factor on the order of two.

To asses the most favorable situation for of a series of HRC (i.e. multiple HRR) between the same bodies, we are interested in the overall fraction of runners that return with an HRC. There are two competing effects: for runner departing at higher velocity, the proportion of HRR being again a HRC increases, but at the same time the fraction of runner that return to the target is reduced. Taking into account both effects, we observe that the overall number of HRC returns increases with the departing velocity, with up to 9\% of the dynamical evolution runs performed in the $m_\mathrm{imp}=\SI{0.2}{\mearth}$ case (both for $v_\mathrm{coll}/v_\mathrm{esc}=1.15$ with $\theta_\mathrm{coll}=\SI{60.0}{\degree}$ and $v_\mathrm{coll}/v_\mathrm{esc}=1.20$ with $\theta_\mathrm{coll}=\SI{52.5}{\degree}$), and up to 7\% in the $m_\mathrm{imp}=\SI{0.5}{\mearth}$ case. Even though unlikely, multiple HRC between the same bodies is possible. This kind of events requires an initial collision that produces a rapidly departing runner since the relative velocity is expected to decrease at each encounter, ultimately ending in a merging collision. Furthermore, the mass contrast is increased following a HRC, as net mass transfer from the impactor to the target is the norm. Hence, a return collision with more dissimilar sized bodies and lower velocity reduces the likelihood of grazing events, either GMC or HRC. The last effect does not strongly affects the results, as the mass change is relatively small, less than 10\% of the impactor's mass usually, which results in change of radius of a few percents at most.

As for more rapidly departing runner the fraction of collisions with other bodies is comparable with the one of the target, another path for multiple HRC is with intermediate collisions occurring with other bodies.

We discussed in Section~\ref{sec:ret-vel} that the presence of additional bodies tend to increase the scatter of the velocity of the return collision. For grazing collisions, the outcome is strongly depending on the velocity in this range. Thus, the results presented here will also depend on the configuration of the system. However, if the impact angle and velocity are uncorrelated, then the effect is lower, as lower-velocity collision will tend to favor GMC while HRC are preferred for more energetic events. For instance, if we compare the types obtained with the obtained from the dynamical evolution with no other bodies present, then we obtain that the types of collisions are quite similar, with in a few series a lower number of HRC.

In our situation, there is mainly one body with which the collision remnants can closely interact: Venus. With more bodies present, such as what can be expected during the formation of planetary systems, the number of interactions would increase, thereby reducing the relationship of the velocity between successive collisions. If that case, the fraction of HRC is also less constrained, making collision chains more probable.

\section{Discussion and conclusion}
\label{sec:discussion}

In this work, we model hit and run collisions (HRC) that are likely to occur at the end of planetary systems formation (the ``late stage'') and follow the dynamical evolution of the resulting bodies to see if they re-collide. In the following section, we summarize the overall conclusions, and discuss the implications of the results on scenarios proposed to explain certain features of the solar system.

\subsection{Return probability}

Under the gravitational influence of other bodies, the return probability of the runner is heavily reduced. This is particularly true for moderately fast runners ($v_\mathrm{dep}/v_\mathrm{esc}\ge1.1$). When the dynamical evolution is performed without other bodies present, that is, only the Sun, the target and the runner, then the return probability is barely affected by the departing velocity, and is always around 90\%. But this is an unrealistic scenario. When the other inner planets of the Solar system are included during the dynamical evolution (up to and including Saturn) then the return probability is greatly affected the departing velocity. Even for moderate departing velocities on the order of $v_\mathrm{dep}/v_\mathrm{esc}\simeq1.1$, the likelihood that the runner collides with a different target, is similar to the likelihood that it returns to the same initial target. In summary, the assumption that runners return to the same target body is valid only when there are no other bodies present. When a system of planets is present, direct modeling of the runner trajectory is necessary to obtain a good fidelity of the results.

\subsection{Delay between successive collisions}

We find delays between successive collisions on the order of \num{e5}--\num{e6} years for collisions occurring at \SI{1}{\au}. We expect that the location of the collision to be important, as the delay should be related to the orbital period. For comparison, \citet{2013IcarusChambers} found that treating collisions more realistically, i.e. by adding the possibility of multiple remnants, increases the formation time by a factor of about two. Our delays are thus consistent with these findings, as they are shorter than the time span for the formation of the solar system's terrestrial planets (a few tens of million of years).

For planets forming during the earlier times of a stellar system, i.e. before the nebula vanishes (around \num{3e6} years; \citealp{2009AIPCMamajek,2016ApJLi}), the consequences can be different. In this case the time span between collisions is comparable to the duration of the stage, except for the close-in planets.

\subsection{Collision geometry}

The orientation between the orbital planes of the prior and returning collisions is essentially compatible with a uniform distribution. The return collision has then no knowledge of the orientation of the prior impact.
If our results are also applicable to any series of giant impacts, then it poses constraints on scenarios that involve multiple collisions, or the implications thereof.

Giant impacts are a convenient mean to explain planetary rotations, and grazing collisions are found useful to explain satellite systems: Earth's Moon \citep{1986IcarusBenz,2001NatureCanup,2004IcarusCanup}, Mars' Phobos and Deimos \citep{2015IcarusCitron,2017ApJHyodo}, Pluto-Charon \citep{2005ScienceCanup,2011AJCanup}, and Haumea \citep{2010ApJLeinhardt}. 
However, if the orientation of giant impacts is not correlated, then it is unlikely for multiple collisions to happen close to the equatorial plane of the target body and in the prograde direction, first to spin up the body into a rapid rotator, and then for the proposed Moon-forming collision to take place. This greatly diminishes the likelihood of scenarios that require such a situation \citep{2008IcarusCanup,2012ScienceCuk}.

If the impactors indeed follow a uniform distributions, scenarios that require alignments of multiple collisions, such as in \citet{2017NatGeoRufu} and \citet{2018ApJCitron}, are also very unlikely. Each collision would contribute to the total angular momentum with a different orientation, and the final value would be random, or close to zero, leaving no possibility for a satellite to survive. Hence, if our Moon is the product of one or more collisions, only a small number of those would be able to provide the rotationally-oriented material from which the Moon accretes.

\subsection{Type of the return collision}

Energy dissipation through shocks during the collision leads to a runner departing at a lower velocity compared to the pre-impact velocity. The steeper the impact angle, the greater the dissipation. Since the median returning velocity is the same a the departing velocity, the return collisions happen usually at a lower velocity than the initial collision. Combining this with the normal slight mass transfer from the impactor to the target during a HRC, and the returning collision has a greater likelihood to be a merger. 

A consequence is that a series of low-velocity HRR between the same objects is unlikely; in order to maintain a sufficiently high impact velocity for the returning collision to also be in the HRC regime, the initial collision must happen at a high velocity. In the latter case however, the probability of the runner returning to the same body is diminished, leaving overall a low-probability for multiple HRR in series with the same target, and a greater probability of scenarios where a higher-energy runner goes from target to target. When considering the ultimate survival of a runner, as proposed by \citep{2014NatGeoAsphaug} for the origin of Mercury, one must therefore consider a longer-term dynamical evolution to determine whether a runner eventually becomes accreted or survives.

To track the runner with sufficient accuracy in general planetary formation simulations, a sufficiently precise collision model is required. This model must not only provide the masses of the resulting bodies, but also their orbital parameters. Furthermore, the outcome for HRC must be properly returned, as it is quite different from a close encounter: relative velocity is reduced and mass transfer occurs.

\subsection{Material mixing}

HRC are known to be able to alter the bulk composition of bodies, mostly by stripping the outer layer, either silicate \citep{2009ApJMarcus} (and supposedly for the specific case of Mercury) or for water ice \citep{2010ApJMarcus,2018CeMDABurger}. Composition changes can produce diverse types of planetesimals, which could potentially explain specific Earth's element content \citep{2015IcarBonsor,2015ApJCarter}.

Collision-induced mixing also has implications in the general context of planetary formation. Relatively head-on HRCs can be a source of significant material mixing, where in typical simulations, about 20\% of the runner's mantle is composed of target's material, while the target's mantle bears nearly 10\% of impactor material. HRC are therefore a potential source of equilibration between the planetary bodies. Coupled with a low return probability on the same body, moderately fast runners have the ability to impact other objects afterwards, leading to a complex picture of geochemical evolution that can provide a source of equilibration between planetary bodies. 

Looking at the collision probabilities from Table~\ref{tab:return-frac}, we note that a runner from the Earth reaching Venus is roughly ten times more likely than Mars. It is therefore possible that Venus and Earth have equilibrated following an HRC on Earth whose runner later impacted Venus while Earth and Mars did not. If this was the case, then Venus would show a composition more similar to Earth than expected from standard formation scenarios.

\acknowledgments

The authors thank Saverio Cambioni, Travis~S.~J. Gabriel, Alan~P. Jackson, and Stephen~R. Schwartz for fruitful discussions, and an anonymous reviewer for his comments and edits which helped improve the clarity of the manuscript. 
We acknowledge support from NASA grant NNX16AI31G and the University of Arizona.
An allocation of computer time from the UA Research Computing High Performance Computing (HPC) is gratefully acknowledged.

\software{Mercury \citep{1999MNRASChambers}, matplotlib \citep{2007HunterMPL}}

\bibliographystyle{aasjournal}
\bibliography{return}

\appendix

\section{Searching for bodies}
\label{sec:bodies}

\begin{figure*}
\centering
\includegraphics{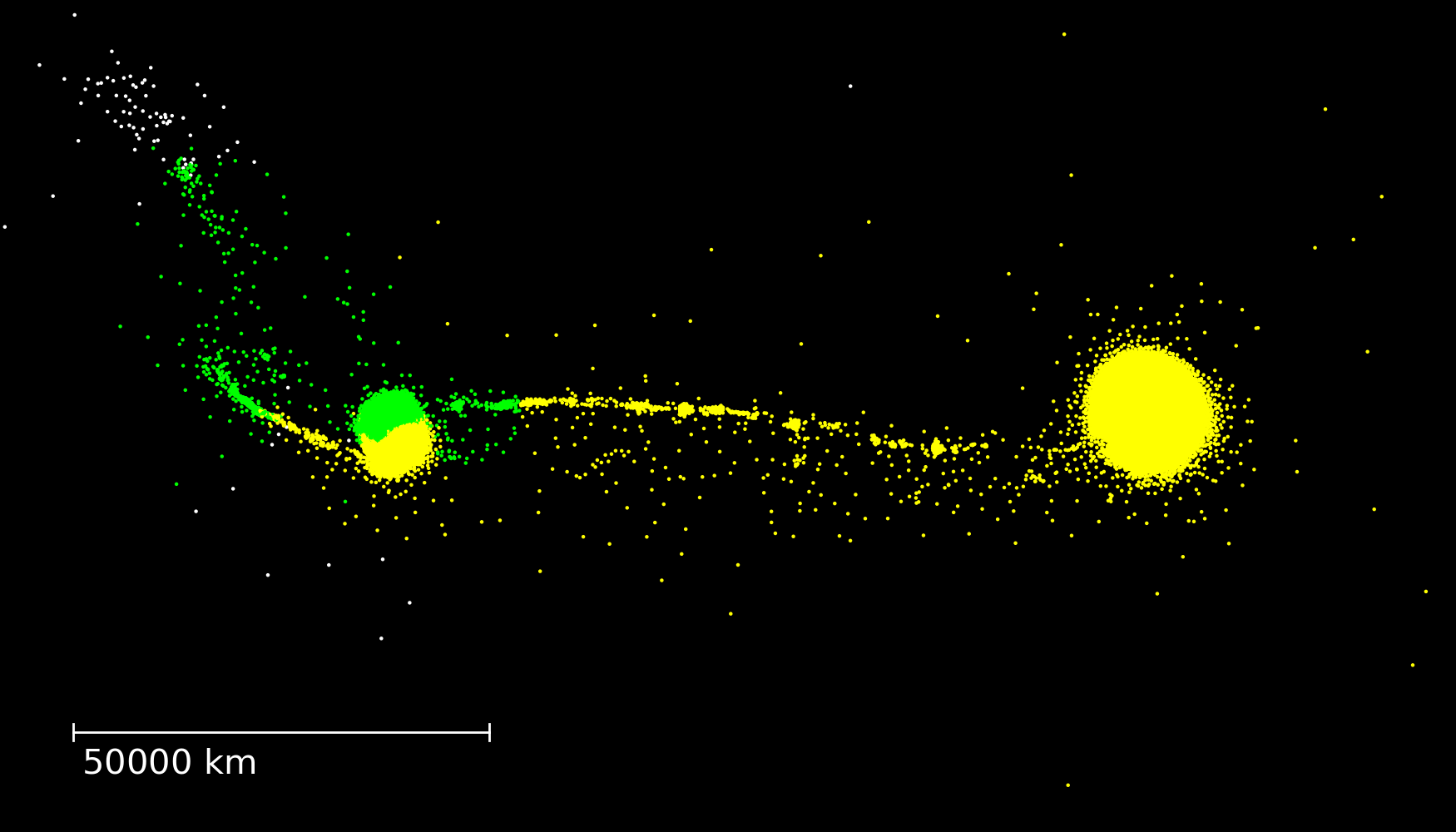}
\includegraphics{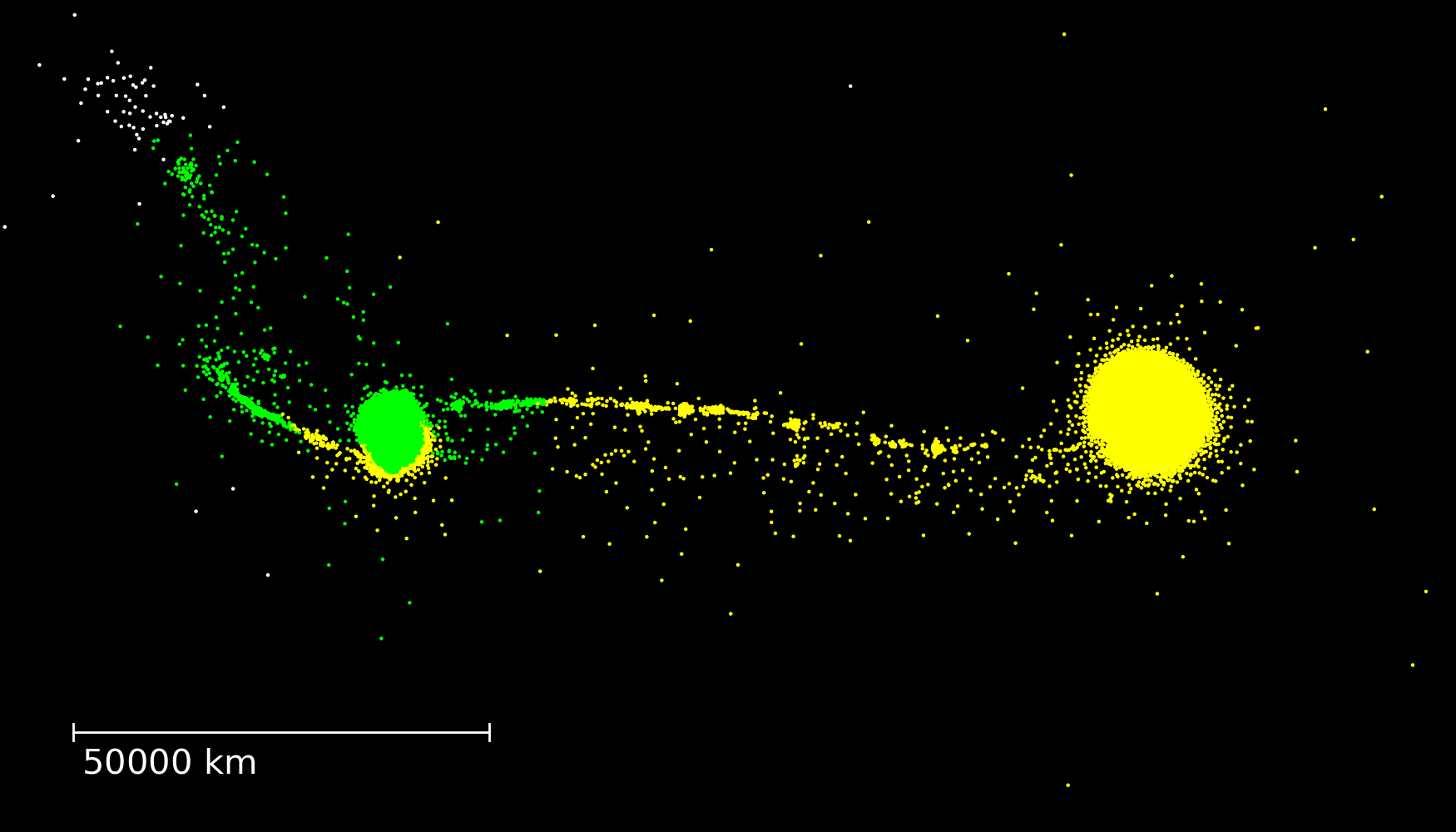}
\caption{State of the collision with $m_\mathrm{imp}=\SI{0.2}{\mearth}$, $v_\mathrm{coll}/v_\mathrm{esc}=1.20$, and $\theta_\mathrm{coll}=\SI{42.5}{\degree}$ at \SI{6}{\hour} after initial contact. The collision is counter-clockwise and the impactor comes from the right. Particles are colored according to the body to which they belong to, with yellow for the largest remnant, green for the second remnant and white for unbound particles. Two different methods are being compared: simple gravity search (\textit{left}) and by stating with a friends-of-friends search followed by gravity (\textit{right}).}
\label{fig:clumps-example}
\end{figure*}

\begin{figure*}
\centering
\includegraphics{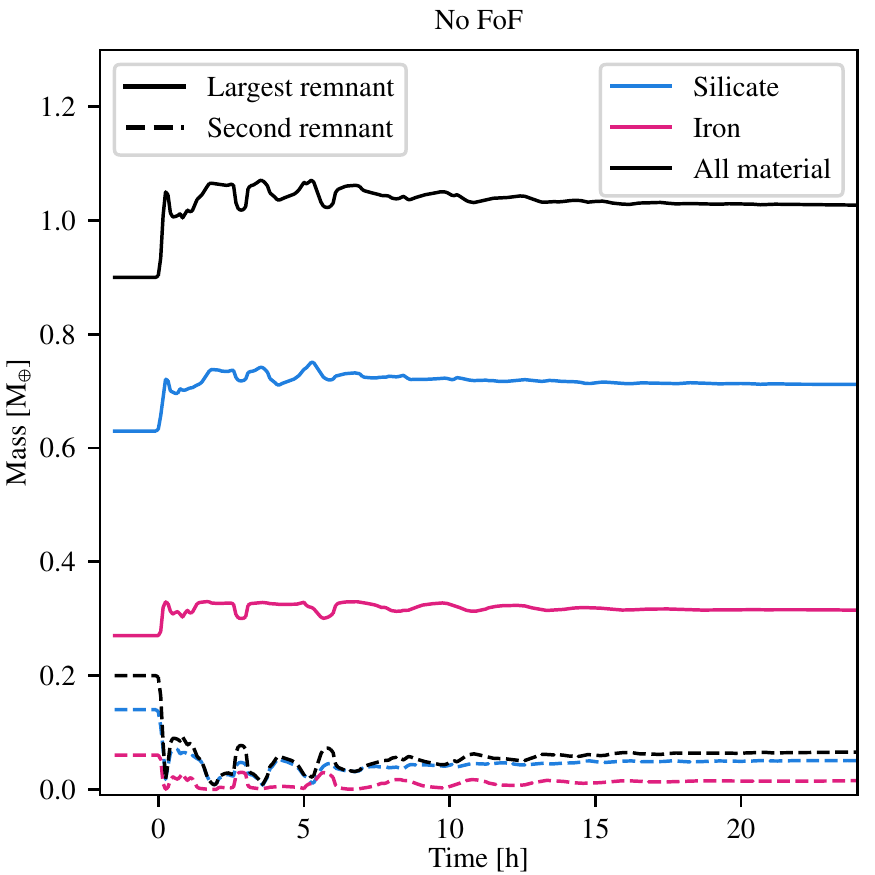}
\includegraphics{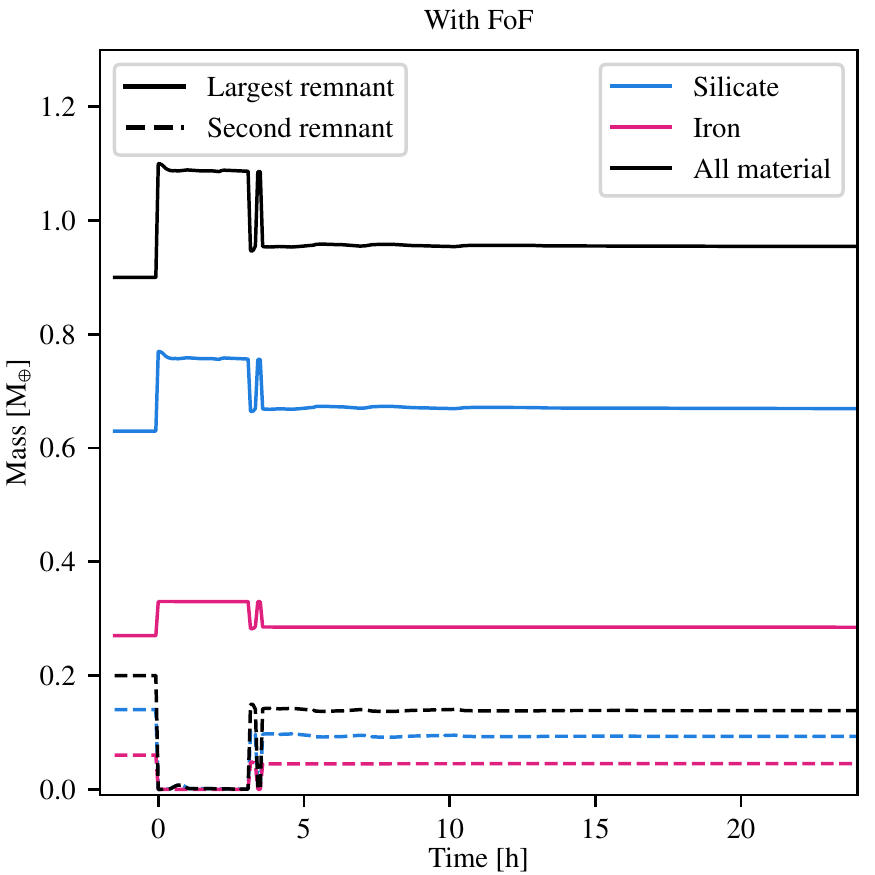}
\caption{Mass of the clumps versus time since initial contact for the collision with $m_\mathrm{imp}=\SI{0.2}{\mearth}$, $v_\mathrm{coll}/v_\mathrm{esc}=1.20$, and $\theta_\mathrm{coll}=\SI{42.5}{\degree}$ for two different methods: simple gravity search (\textit{left}) and by starting with a friends-of-friends search followed by gravity (\textit{right}). The values at negative times give the initial state.}
\label{fig:clumps-mass}
\end{figure*}

Searching for remnants in GMC or HRC close to the boundary between the two regimes can be tricky. Small fluctuations in velocities are sufficient to result in the bodies being gravitationally bound or not. For runners that are barely escaping or barely bound, the problem is exacerbated by the rotation induced from the encounter. A simple gravitational search, treating all SPH particles independently, fails to grasp this effect, and treats parts of the rotating runner in different ways. An example of such behaviour is shown in the left panel of Fig.~\ref{fig:clumps-example}. It can be seen that a part of the runner, while unbound as a whole, is still found to be part of the largest remnant. The collision happens counter-clockwise, and so does the spin of the bodies. Particles whose motion due to body-rotation is towards the target (in the bottom in Fig.~\ref{fig:clumps-example}) are found to be bound to the target rather than the runner.

Starting the search for bodies by a friends-of-friends (FoF) walk mitigates most of this problem and is the approach we have taken here. For this search, particles that have a density of at least 3/4 of the reference density from the equation of state and are closer than twice their mean smoothing length are deemed to form a single body. For the later gravity search, bodies determined by the FoF method are treated as a single super-particle. With this scheme, the runner is first determined as a single body and remains as such during the gravity search. Hence the rigid body rotation is averaged out and a negligible influence on the search algorithm. Only a small number of particles lying on the surface of the runner are not found by the FoF algorithm and are still deemed to be part of the largest remnant, but their contribution to the total mass is small.

The final outcome of the two methods is shown in Fig.~\ref{fig:clumps-mass}. With a simple gravity search, the clump masses take about \SI{18}{\hour} to converge while when a FoF search is included the whole set is initially found to be a single body until the arm that connects them vanishes and then only small fluctuations continues. But most importantly, the two methods do not converge to the same value; the simple gravity search misses a part of the runner that happens to be rotating towards the target. In addition, since the part of the runner that is found to be part of the largest body has a net motion, other properties such as the relative velocity are affected. The converse in a GMC, where part of the transient body is found to not be bound to the largest body, is also possible. Hence a FoF search is a necessity for outcomes of collisions that are close to the transition between GMC and HRC regimes. 

FoF search has a further advantage in GMC. Gravity search is unable to distinguish the transient body, as it is still bound to the target. FoF is not affected by this, and makes us able to compute the orbit of the transient body until later accretion. We will use this feature to more accurately evolve weakly-bound GMC where the body has still not been accreted after 24 hours.  

\end{document}